\documentclass[journal]{IEEEtran}

\usepackage{xcolor,soul,framed} 

\colorlet{shadecolor}{yellow}
\usepackage[pdftex]{graphicx}
\graphicspath{{../pdf/}{../jpeg/}}
\DeclareGraphicsExtensions{.pdf,.jpeg,.png}
\usepackage{amsfonts}
\usepackage{multirow}

\usepackage[utf8x]{inputenc}

\usepackage[cmex10]{amsmath}
\usepackage{amssymb}
\usepackage{array}
\usepackage{mdwmath}
\usepackage{mdwtab}
\usepackage{eqparbox}
\usepackage{url,balance,comment}
\usepackage[noadjust]{cite}
\usepackage{tikz}
\usepackage{tikzscale}
\usepackage{graphicx}
\usepackage{tkz-euclide} 
\usetikzlibrary{shapes,arrows} 
\usepackage{tikz-dimline}
\usetikzlibrary{arrows.meta}
\usetikzlibrary{external}
\ifCLASSOPTIONcompsoc
\usepackage[caption=false,font=normalsize,labelfon
t=sf,textfont=sf,subrefformat=parens,labelformat=parens]{subfig}
\else
\usepackage[caption=false,font=footnotesize,subrefformat=parens,labelformat=parens]{subfig}
\fi
\usepackage{url}

\usetikzlibrary{arrows}
\usetikzlibrary{automata,positioning}
\usetikzlibrary{patterns}
\usepackage[ruled,vlined,linesnumbered]{algorithm2e}

\usepackage{color}

\usepackage{mathtools}
\usepackage[super]{nth}
\newlength{\hsep}

\usepackage[noadjust]{cite}
\SetKwInput{KwData}{Input}
\SetKwInput{KwResult}{Output}

\newcommand{\Tzeta}{\Tilde{\zeta}}
\newcommand{\RP}{\text{RP}}
\newcommand{\LP}{\text{LP}}
\newcommand{\FLP}{\text{FLP}}
\newcommand{\Azg}{A_0^{(\gamma)}}
\newcommand{\Aog}{A_1^{(\gamma)}}
\newcommand{\Azb}{A_0^{(\beta_2)}}
\newcommand{\Aob}{A_1^{(\beta_2)}}
\newcommand{\Azt}{A_0^{(\theta)}}
\newcommand{\Aot}{A_1^{(\theta)}}

\usepackage{pgfplots}
\pgfplotsset{compat = newest}

\newcommand{\TA}{\Tilde{A}}

\newcommand{\argwz}{(\omega,z)}
\newcommand{\argtz}{(t,z)}

\DeclareMathOperator*{\argmax}{\text{argmax}} 

\def\crosssize{0.065in}
\tikzset{cross/.style={cross out, draw, minimum size=\crosssize,thick, inner sep=0pt, outer sep=0pt}}

\graphicspath{{figures/}}

\begin{document}
\bstctlcite{IEEEexample:BSTcontrol}
    \title{Frequency Logarithmic Perturbation on the Group-Velocity Dispersion Parameter with Applications to Passive Optical Networks}
  \author{Vinícius~Oliari,~\IEEEmembership{Student Member,~IEEE,}
      Erik~Agrell,~\IEEEmembership{Fellow,~IEEE}, Gabriele~Liga,~\IEEEmembership{Member,~IEEE}, and Alex~Alvarado,~\IEEEmembership{Senior Member,~IEEE}
  \thanks{
    V. Oliari, G. Liga, and A. Alvarado are with the Signal Processing Systems (SPS) Group, Department of Electrical Engineering, Eindhoven University of Technology, 5600 MB Eindhoven, The Netherlands (e-mails: v.oliari.couto.dias@tue.nl, a.alvarado@tue.nl).
  
    E. Agrell is with the Department of Electrical Engineering, Chalmers University of Technology, Gothenburg SE-41296, Sweden (e-mail: agrell@chalmers.se).
  
    This work is supported by the Netherlands Organisation for Scientific Research (NWO) via the VIDI Grant ICONIC (project number 15685). The work of A. Alvarado has received funding from the European Research Council (ERC) under the European Union's Horizon 2020 research and innovation programme (grant agreement No 757791). The work of E. Agrell has received funding from the Swedish Research Council (VR) under Grant no. 2017-03702. The work of G.~Liga is supported by the EuroTechPostdoc programme under the European Union's Horizon 2020 research and innovation programme (Marie Sk\l{}odowska-Curie grant agreement No 754462).}
}  


\maketitle

\begin{abstract}
Signal propagation in an optical fiber can be described by the nonlinear Schrödinger equation (NLSE). The NLSE has no known closed-form solution, mostly due to the interaction of dispersion and nonlinearities. In this paper, we present a novel closed-form approximate model for the nonlinear optical channel, with applications to passive optical networks. The proposed model is derived using logarithmic perturbation in the frequency domain on the group-velocity dispersion (GVD) parameter of the NLSE. The model can be seen as an improvement of the recently proposed regular perturbation (RP) on the GVD parameter. RP and logarithmic perturbation (LP) on the nonlinear coefficient have already been studied in the literature, and are hereby compared with RP on the GVD parameter and the proposed LP model. As an application of the model, we focus on passive optical networks. For a 20 km PON at 10 Gbaud, the proposed model improves upon LP on the nonlinear coefficient by 1.5 dB. For the same system, a detector based on the proposed LP model reduces the uncoded bit-error-rate by up to 5.4 times at the same input power or reduces the input power by 0.4 dB at the same information rate.

\end{abstract}

\begin{IEEEkeywords}
Channel modeling, chromatic dispersion, Kerr nonlinearity, logarithmic perturbation, nonlinear Schrödinger equation, optical fiber, regular perturbation, weakly dispersive regime.
\end{IEEEkeywords}

\IEEEpeerreviewmaketitle

\section{Introduction}

\IEEEPARstart{A}{nalytical} models for optical fiber transmission have been widely studied in the literature. These models are based on the equations that govern the optical field propagation: the nonlinear Schr\"odinger equation (NLSE) \cite[Ch.\ 2]{AGRAWAL201327} and its variants. The NLSE has no known exact solution for an arbitrary input waveform. One of the most efficient alternatives for approximated numerical solutions is the split step Fourier method (SSFM) \cite{Sinkin2002}, which simulates the effects of fiber propagation. On the other hand, to analyse these effects and design novel transceivers, analytical models are highly desirable.

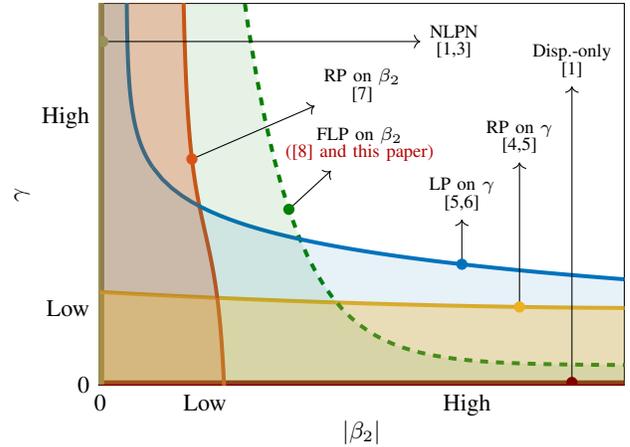
\begin{figure}[t!]
\centering
\begin{tikzpicture}

\definecolor{mycolor2}{rgb}{0.00000,0.44706,0.74118}%
\definecolor{mycolor1}{rgb}{0.00000,0.49804,0.00000}%
\definecolor{mycolor3}{rgb}{0.92900,0.69400,0.12500}%
\definecolor{mycolor4}{rgb}{0.85098,0.32549,0.09804}%
\definecolor{mycolor5}{rgb}{0.39,0.19,0.0625};
    
\def\slo{\small}
\def\to{\footnotesize}

\begin{axis}[
width=2.75in,
height=2in,
at={(0.09in,1.875in)},
scale only axis,
xmin=-0.025,
xmax=10,
ymin=-0.025,
ymax=10,
tick style={draw=none},
axis line style={line width=0.35pt},
yticklabels={{0}, {Low}, {High}},
ytick = {0,2,7},
xticklabels={{0}, {Low}, {High}},
xtick = {0,2,7},
xlabel = {$|\beta_2|$},
ylabel = {$\gamma$},
ylabel style={font=\color{white!15!black},font=\slo},
xlabel shift = -0.07in,
ylabel shift = -0.03in,
]
\def\lw{1.5pt};
\def\arrsz{0.35};
\def\stnlpn{1.15};
\def\stdisponly{1.75};
\def\rpgammaH{0.85};

\tikzstyle{every node}=[font=\small]

\addplot [
    domain=0:10, 
    samples=5,
    color=red!50!black,
    line width = 2*\lw,
]
{0};

\addplot [
    domain=0:0.01, 
    samples=5,
    color=yellow!50!black,
    line width = 2*\lw,
]
{10000*x};

\addplot [
    domain=0:10, 
    samples=15, 
    color=mycolor3,
    line width = \lw,
]
{0.0035*(x-11)*(x-11)+2};

\addplot [
    domain=1.5:2.365, 
    samples=50, 
    color=mycolor4,
    line width = \lw,
]
{-2*tan(175*(x-2))+4};

\addplot [
    domain=2:10, 
    samples=50, 
    color=mycolor1,
    line width = \lw,
    dashed
]
{0.0025*exp(-x+11)+0.5};

\addplot [
    domain=0.5:1, 
    samples=200, 
    color=mycolor2,
    line width = \lw
]
{-ln((x-0.5)/0.0025)+11};

\addplot [
    domain=1:10, 
    samples=50, 
    color=mycolor2,
    line width = \lw,
]
{-ln((x-0.5)/0.0025)+11};

\def\stepH{0.75};

\fill [mycolor1, fill opacity=0.1, domain= 2:10, variable=\xx,samples=50]
(0,0) -- (
0,10) -- 
plot ({\xx}, {0.0025*exp(-\xx+11)+0.5}) -- (10,0) -- cycle
;

\fill [mycolor4, fill opacity=0.3, domain=1.5:2.365, variable=\x]
(0,0) -- (0,10) -- 
plot ({\x}, {-2*tan(175*(\x-2))+4})
-- (2.365,0) -- cycle;
;

\fill [mycolor2, fill opacity=0.1, domain= 0.505:10, variable=\xx,samples=300]
(0,0) -- (
0,10) -- 
plot ({\xx}, {-ln((\xx-0.5)/0.0025)+11}) -- (10,0) -- cycle
;

\fill [mycolor3, fill opacity=0.3, domain= 0:10, variable=\xx,samples=50]
(0,0) -- (
0,10) -- 
plot ({\xx}, {0.0035*(\xx-11)*(\xx-11)+2}) -- (10,0) -- cycle
;

\node[align=center] (nln) at (6.75,9) {\scriptsize{NLPN}\\ [-0.85ex]  \scriptsize{[1,3]}};
\draw[->] (0,9)--(nln);
\node [inner sep=1.5,circle,fill=yellow!50!black] at (0.05,9){};

\node[align=center] (rpb2) at (5,7.85) {\scriptsize{RP on $\beta_2$} \\ [-0.85ex]  \scriptsize{[7]}};

\def\xcoord{1.75}
\pgfmathsetmacro{\ycoord}{-2*tan(175*(\xcoord-2))+4}

\draw[->] (\xcoord,\ycoord)--(rpb2);
\node [inner sep=1.5,circle,fill=mycolor4] at (\xcoord,\ycoord){};

\node[align=center] (rpg) at (8,6.5) {\scriptsize{RP on $\gamma$} \\ [-0.85ex]  \scriptsize{[4,5]}};

\def\xcoordg{8}
\pgfmathsetmacro{\ycoordg}{0.0035*(\xcoordg-11)*(\xcoordg-11)+2}

\draw[->] (\xcoordg,\ycoordg)--(rpg);
\node [inner sep=1.5,circle,fill=mycolor3] at (\xcoordg,\ycoordg){};

\node[align=center] (dispo) at (9,8.5) {\scriptsize{Disp.-only} \\ [-0.85ex]  \scriptsize{[1]}};
\draw[->] (9,0)--(dispo);
\node [inner sep=1.5,circle,fill=red!50!black] at (9,0.05){};

\node[align=center] (lpg) at (6.9,5) {\scriptsize{LP on $\gamma$} \\ [-0.85ex]  \scriptsize{[5,6]}};

\def\xcoordlg{6.9}
\pgfmathsetmacro{\ycoordlg}{-ln((\xcoordlg-0.5)/0.0025)+11}

\draw[->] (\xcoordlg,\ycoordlg)--(lpg);
\node [inner sep=1.5,circle,fill=mycolor2] at (\xcoordlg,\ycoordlg){};

\node[align=center] (flpb) at (4.95,6.3) {\scriptsize{FLP on $\beta_2$} \\ [-0.85ex]  \scriptsize{\textcolor{red!70!black}{([8] and this paper)}}};

\def\xcoordflb{3.6}
\pgfmathsetmacro{\ycoordflb}{0.0025*exp(-\xcoordflb+11)+0.5}

\draw[->] (\xcoordflb,\ycoordflb)--(flpb);
\node [inner sep=1.5,circle,fill=mycolor1] at (\xcoordflb,\ycoordflb){};

\end{axis}
\end{tikzpicture}
\caption{Regions in the $\beta_2$ vs $\gamma$ plane where different models are valid. This paper presents a new model for the region to the left of and under the green curve.}
\label{fig:regmod}
\end{figure}

 Some of the NLSE's most used analytical models are only valid under some restricted values of two fiber parameters: the (Kerr) nonlinear coefficient $\gamma$ and the group-velocity dispersion (GVD) parameter $\beta_2$. The validity\footnote{The validity of a model is defined in this paper as the set of parameters values in which the model waveform is sufficiently close to the true waveform obtained from the NLSE. Details on the metric that measures the distance between the two waveforms are given in Sec.~\ref{sc:setup}.} of the models with respect to these parameters is shown in Fig.~\ref{fig:regmod}. When $\gamma$ is equal to zero, the NLSE admits an analytical solution, given by the dispersion-only model \cite[Ch.\ 3]{AGRAWAL201327} and represented by the horizontal line at $\gamma=0$ in Fig.~\ref{fig:regmod}. When $\beta_2$ is set to zero, the NLSE also admits an analytical solution, given by the nonlinear phase noise model (NLPN) \cite[Ch.\ 4]{AGRAWAL201327}, \cite{Beygi:11}, represented by the vertical line at $\beta_2=0$ in Fig.~\ref{fig:regmod}. If both parameters are nonzero, usually a perturbation approach is used \cite{Forestieri2005}.
 
A perturbation on the nonlinear coefficient $\gamma$ considers the nonlinearities a minor effect \cite{Vannucci2002, Forestieri2005} and is accurate for high accumulated dispersion. For the opposite regime, where the nonlinearities are the major effect and the dispersion is low but nonzero, a perturbation on the GVD parameter can be performed. The perturbation techniques usually considered for the optical fiber propagation are regular perturbation (RP) and logarithmic perturbation (LP) \cite{Forestieri2005}. It was shown in \cite{Ciaramella2005} that LP converges faster to the true NLSE's solution than RP. RP and LP on $\gamma$ cover the area under the yellow and blue curves in Fig.~\ref{fig:regmod}, respectively. Recently, we proposed RP on $\beta_2$ in \cite{Oliari2020} and compared with RP on $\gamma$. RP on $\beta_2$ provided more accuracy in the weakly dispersive and highly nonlinear regimes, represented by the area under the red curve in Fig.~\ref{fig:regmod}. A preliminary investigation of LP on $\beta_2$ was reported in our recent work \cite{Oliari2020ecoc}, where LP methods on both $\gamma$ and $\beta_2$ were compared on the waveform level. This paper is an extension of \cite{Oliari2020ecoc}, where our first contribution is to derive the two perturbative models on $\beta_2$ presented in \cite{Oliari2020ecoc}. These models are obtained applying LP in either time or frequency. The latter approach, which we call frequency logarithmic perturbation (FLP), is the most accurate of the two for the $\beta_2$ expansion.


The FLP on $\beta_2$ covers the area under the green dashed curve in Fig.~\ref{fig:regmod}, which improves upon RP on $\beta_2$. FLP should not be confused with the frequency resolved logarithmic perturbation (FRLP) proposed in \cite{SecondiniCM2012,SecondiniAI2013,SecondiniNMWDM2019}. FRLP consists in applying LP on the frequency components of the time-domain signal individually and using the superposition principle to combine them. However, FRLP still applies the LP principle in the time domain, whereas FLP we study here applies this principle in the frequency domain. In this paper, the FLP on $\beta_2$ model is compared with the LP on $\gamma$, RP on $\gamma$, and RP on $\beta_2$ models. 

The second contribution of this paper is to demonstrate the applicability of the proposed FLP on $\beta_2$ model. To this end, we consider passive optical networks (PONs), wherein the accumulated dispersion is typically low. We use the models to estimate optimum decision regions at the receiver. The method used to obtain the decision regions is compared with the Parzen window (PW) method, recently proposed in \cite{karim2019} for optical fiber systems. Simulations are carried out in the C- and O-band, where the latter has a close-to-zero GVD parameter $\beta_2$. At a waveform level (continuous-time), the proposed FLP on $\beta_2$ is shown to outperform the other three models at powers higher than $7$ dBm for both C- and O-band systems. At a symbol level (discrete-time), decision regions obtained from LP on $\beta_2$ result in bit-error-rates (BER) more than five times lower than the ones obtained from LP on $\gamma$. Finally, these decision regions are analysed in a system with forward-error-correction (FEC).

This paper is organized as follows: Section~\ref{sc:prel} provides the mathematical background for the derivation of the models; Section~\ref{sc:setup} compares the models in both continuous- and discrete-time; and Section \ref{sc:conc} concludes the paper.

\section{Mathematical Background} \label{sc:prel}

The normalized NLSE for noiseless propagation of an optical field $A$ at a retarded time frame $t$ and distance $z$ for a single-polarization can be approximated as \cite{Vannucci2002}
 \begin{align}\label{eq:normnlse}
   \dfrac{\partial A(t,z)}{\partial z} = &-\dfrac{j\beta_2}{2}\dfrac{\partial^2 A(t,z)}{\partial t^2 } \notag \\ &+ j\gamma \mathrm{e}^{-\alpha z}|A(t,z)|^2 A(t,z),
 \end{align}
where $\alpha$ is the attenuation coefficient, $\beta_2$ the GVD parameter, and $\gamma$ the nonlinear coefficient. The first term on the right-hand side of \eqref{eq:normnlse} represents the chromatic dispersion. This effect on the waveform $A$ is larger when the bandwidth and/or the fiber length is increased. The last term represents the Kerr nonlinearity, which has a cubic dependence on the instantaneous signal power and also increases with the fiber length. The solution $A$ of \eqref{eq:normnlse} can be numerically estimated by the SSFM \cite[Ch.\ 2]{AGRAWAL201327}, \cite{Sinkin2002}. Other effects, such as third-order dispersion (TOD) \cite[Ch.\ 3]{AGRAWAL201327}, are not taken into account in \eqref{eq:normnlse}. TOD becomes significant for large bandwidths or when $\beta_2$ is low. This effect will be taken into account for our simulations in the O-band, although it is not used in the derivation of the models. For the C-band, we consider solely the effects in \eqref{eq:normnlse}.  

In what follows, we first review three models available in the literature. Sec.~\ref{sc:rpg} and Sec.~\ref{sc:rpb} describe the RP on $\gamma$ and on $\beta_2$, respectively, while Sec.~\ref{sc:lpg} presents the LP on $\gamma$. Finally, Sec.~\ref{sc:flp} introduces the FLP on $\beta_2$.


\subsection{Regular Perturbation on the Nonlinear Coefficient}\label{sc:rpg}

The RP on $\gamma$ was first derived in \cite{Vannucci2002,Forestieri2005}. To approximate the solution of \eqref{eq:normnlse}, the RP method represents the signal by a power series of a certain coefficient. For the RP on $\gamma$, the RP solution can be written as

\begin{align}\label{eq:gammarp}
    A(t,z) &= \sum_{k=0}^{\infty} \gamma^k A_{k}^{(\gamma)} (t,z).
\end{align}
To obtain the functions $A_k^{(\gamma)}$, \eqref{eq:gammarp} is substituted into \eqref{eq:normnlse} and the terms multiplied by the $k$-th power of $\gamma$ are equated. An approximate solution for \eqref{eq:normnlse} can be obtained by considering only the functions $A_0^{(\gamma)}$ and $A_1^{(\gamma)}$. This approximation is the first-order RP on $\gamma$ \cite[Eqs.\ (7), (9)]{Vannucci2002}, \cite[Eq.\ (12)]{Forestieri2005}

\begin{equation}\label{eq:rpg1}
    A(t,z) \approx A_{\RP}^{(\gamma)}(t,z) = A_0^{(\gamma)}(t,z) + \gamma A_1^{(\gamma)}(t,z),
\end{equation}
where
\begin{align}\label{eq:A0gammaRP}
   \Azg(t,z) &=  \mathcal{D}_z \{A(\cdot,0)\}(t), \\
   \Aog(t,z) &= j \int_0^z \! \mathrm{e}^{-\alpha u}\mathcal{D}_{z-u} \left\{ |A_0(\cdot,u)|^2A_0(\cdot,u) \right\} \! (t) \text{d}u,\label{eq:A1gammaRP}
\end{align}
and the dispersion operator $\mathcal{D}_z$ is
\begin{align}
        \mathcal{D}_z \{ f \}(t) &\triangleq \left(f \ast h(\cdot, z) \right)(t), \\
            h(t,z) &= \dfrac{1}{\sqrt{j2\pi \beta_2 z}} \mathrm{e}^{-\frac{j}{2\beta_2 z}t^2}.
\end{align}
The function $\Azg$ in \eqref{eq:A0gammaRP} is called the dispersion-only solution of \eqref{eq:normnlse}. This solution can be seen as a model that is accurate only when the nonlinear effect is negligible. The first-order RP on $\gamma$ in \eqref{eq:rpg1} is accurate for low nonlinear effects and is illustrated as the yellow curve in Fig.~\ref{fig:regmod}.

\subsection{Regular Perturbation on the GVD Parameter}\label{sc:rpb}

We recently proposed the RP on $\beta_2$ in \cite{Oliari2020}. The same procedure as in \eqref{eq:gammarp} can be applied by considering $A$ as a power series of $\beta_2$, i.e.,

\begin{equation}\label{eq:genrpb2}
     A(t,z) = \sum_{k=0}^{\infty} \beta_2^k A_{k}^{(\beta_2)}(t,z).
\end{equation}
In analogy to RP on $\gamma$, the functions $A_k^{(\beta_2)}$ are also obtained by replacing \eqref{eq:genrpb2} in \eqref{eq:normnlse} and equating the terms related to the $k$-th power of $\beta_2$. For the first-order RP, involving the functions $\Azb$ and $\Aob$, an approximate solution for $A$ can be obtained by

\begin{equation}\label{eq:rpb2_1}
    A(t,z) \approx A_{\RP}^{ (\beta_2)}(t,z) = \Azb(t,z) + \beta_2 \Aob(t,z),
\end{equation}
where 
\begin{equation}\label{eq:A0b2RP}
     \Azb(t,z) = A(t,0)\mathrm{e}^{j\gamma |A(t,0)|^2 G(z)},
\end{equation}
and
\begin{equation}\label{eq:A1b2RP}
    \Aob(t,z) = B(t,z)\mathrm{e}^{j \gamma |A(t,0)|^2 G(z)},
\end{equation}    
with $B$ given by
\begin{align}\label{eq:Bb2th}
  B(t,z) = &-M(t)z+ G_{1}(z) R(t)+G_{2}(z)P(t) \notag \\ &-2j\gamma A(t,0) \Re\{A^*(t,0) V(t,z)\} ,\\
       V(t,z)= & \  G(z) \left[M(t)z- G_{1}(z) R(t)-G_{2}(z)P(t) \right]  \notag \\
      &   -G_{1}(z)M(t) +G_{2}(z)R(t)+G_{3}(z)P(t) ,\\
    M(t) &= \dfrac{j}{2}\dfrac{\partial^2 A(t,0)}{\partial t^2}, \label{M}\\
    R(t) &= \dfrac{\gamma }{2} A(t,0)\dfrac{\partial^2 |A(t,0)|^2}{\partial t^2}+   \gamma \dfrac{\partial A(t,0)}{\partial t} \dfrac{\partial |A(t,0)|^2}{\partial t},\label{R}\\
    P(t) &= \dfrac{j \gamma^2}{2}  A(t,0)  \left(\dfrac{\partial |A(t,0)|^2}{\partial t}\right)^2,\label{P}\\
    G_{1}(z) &= \dfrac{\alpha z+\mathrm{e}^{-\alpha z}-1}{\alpha^2},\label{eq:g1} \\
    G_{2}(z) &= \dfrac{ 2\alpha z+4\mathrm{e}^{-\alpha z}-\mathrm{e}^{-2\alpha z}-3}{2\alpha^3},\label{eq:g2}\\
    G_{3}(z) &= \dfrac{ 6 \alpha z+18 \mathrm{e}^{-\alpha z}-9 \mathrm{e}^{-2 \alpha z}+2\mathrm{e}^{-3\alpha z}-11}{6 \alpha^4}.\label{eq:g3}
\end{align}
Analogously to RP on $\gamma$, the function $\Azb$ in \eqref{eq:A0b2RP} for RP on $\beta_2$ is an accurate model when dispersion is negligible, and is called the NLPN model \cite{Beygi:11,Oliari2020}. The first-order RP on $\beta_2$ in \eqref{eq:rpb2_1} is accurate for low accumulated dispersion and is illustrated as the red curve in Fig.~\ref{fig:regmod}.


\subsection{Logarithmic Perturbation}\label{sc:lpg}

LP is a mathematical technique similar to RP. LP on $\gamma$ was first presented in \cite{Ciaramella2005,Forestieri2005} and can be shown to converge faster to the true NLSE's solution than RP on $\gamma$. LP functions can be obtained directly by the RP functions $A_k$. For example, following an approach similar to \cite{Forestieri2005}, consider that the signal $A$ can be written as a power series of a coefficient $\theta$ (e.g., $\gamma$ or $\beta_2$ as done in \eqref{eq:gammarp} and \eqref{eq:genrpb2}) as
\begin{equation}\label{eq:genrp}
    A \argtz = \sum_{k=0}^{\infty}\theta^k A_k^{(\theta)} \argtz,
\end{equation}
where $A_k^{(\theta)}$ is the $k$-th RP function. We now want to express $A$ in its LP version, which takes the form
\begin{equation}\label{eq:genlp}
    A \argtz  = A_0^{(\theta)} \argtz \exp\left(\sum_{k=1}^\infty \theta^k \psi_k^{(\theta)} \argtz \right),
\end{equation}
where the function $\psi_k^{(\theta)}$ is the $k$-th LP function and $A_0^{(\theta)}$ is the $0$-th order RP function. Representing the exponential function in \eqref{eq:genlp} by its Taylor expansion yields 
\begin{equation}
 \label{eq:genlptay}
     A \argtz = A_0^{(\theta)} \argtz \sum_{m=0}^\infty  \dfrac{1}{m!} \left(\sum_{k=1}^{\infty} \theta^k \psi_k^{(\theta)}  \argtz \right)^m .  
\end{equation}

The functions $\psi_k^{(\theta)} $ can now be obtained by equating \eqref{eq:genlptay} with \eqref{eq:genrp}, and further equating the terms that have the same power of $\theta$. For example, equating the terms multiplied by $\theta^1$, we obtain the first-order LP function as
\begin{equation}\label{eq:genlp1}
    \psi_1^{(\theta)}  \argtz = \dfrac{\Aot \argtz}{\Azt \argtz}.
\end{equation}
Equating the terms multiplied by $\theta^2$ we can also obtain the second-order LP function
\begin{equation}\label{eq:genlp2}
    \psi_2^{(\theta)}  \argtz =  \dfrac{A_2^{(\theta)} \argtz}{\Azt \argtz}-\dfrac{1}{2}\left(\dfrac{\Aot \argtz}{\Azt \argtz}\right)^2.
\end{equation}
The function $\psi_2^{(\theta)}$ in \eqref{eq:genlp2} depends on the RP term $A_2^{(\theta)}$. For RP on $\gamma$, $A_2^{(\gamma)}$ is well defined \cite[Eq.\ (11)]{Vannucci2002}, \cite[Eq.\ (12)]{Forestieri2005}. However, for RP on $\beta_2$, $A_2^{(\beta_2)}$ is not known in the literature at the time this paper is being written. Thus, we will restrict the analysis to first-order LP and RP only.

\begin{table*}
\centering
\caption{Summary of the first-order perturbation methods discussed in this paper}
\begin{tabular}{ |c||c|c|c|c|c|c|  }
 \hline
 \multirow{4}{*}{Coeff. $\theta$} & \multicolumn{6}{c|}{Perturbation method} \\
 \cline{2-7}
 & \multicolumn{2}{c|}{\nth{1} order RP} & \multicolumn{2}{c|}{\nth{1} order LP} &  \multicolumn{2}{c|}{\nth{1} order FLP} \\
 & \multicolumn{2}{c|}{$A(t,z) = A_0^{(\theta)}\argtz+\theta A_1^{(\theta)}\argtz$} & \multicolumn{2}{c|}{$A\argtz  = A_0^{(\theta)}\argtz\exp{\left(\theta\psi_1^{(\theta)}\argtz\right)} $} & \multicolumn{2}{c|}{$\TA\argwz  = \TA_0^{(\theta)}\argwz\exp{\left(\theta\Tzeta_1^{(\theta)}\argwz\right)} $}
\\
\hline
\multirow{4}{*}{$\gamma$} & \multirow{2}{*}{$A_0^{(\gamma)}\argtz$} & \multirow{2}{*}{$ \mathcal{D}_z \{A(\cdot,0)\}(t)$} &  \multirow{4}{*}{$\psi_1^{(\gamma)} \argtz$}  & \multirow{4}{*}{$\dfrac{\Aog \argtz}{ \Azg \argtz}$} & 
\multirow{4}{*}{$\Tzeta_1^{(\gamma)} \argwz$ }& \multirow{4}{*}{$\dfrac{\TA_1^{(\gamma)} \argwz}{\TA_0^{(\gamma)} \argwz}$ }  \\ & & & & & & \\
\cline{2-3} 
&  \multirow{2}{*}{$\Aog\argtz$} &  \multirow{2}{*}{ \eqref{eq:A1gammaRP}} &  &  & & \\   & & & & & & \\
\hline
\multirow{4}{*}{$\beta_2$} & \multirow{2}{*}{$A_0^{(\beta_2)}\argtz$} & \multirow{2}{*}{$ A(t,0)\mathrm{e}^{j\gamma |A(t,0)|^2 G(z)}$} &  \multirow{4}{*}{$\psi_1^{(\beta_2)} \argtz$}  & \multirow{4}{*}{$\dfrac{\Aob \argtz}{ \Azb \argtz}$} & 
\multirow{4}{*}{$\Tzeta_1^{(\beta_2)} \argwz$}& \multirow{4}{*}{$\dfrac{\TA_1^{(\beta_2)} \argwz}{\TA_0^{(\beta_2)} \argwz}$ }  \\ & & & & & & \\
\cline{2-3} 
&  \multirow{2}{*}{$\Aob\argtz$} &  \multirow{2}{*}{ \eqref{eq:A1b2RP}} & &  & &   \\ & & & & & & \\
\hline
\end{tabular}
\label{tab:pfunc}
\end{table*}

Setting $\theta=\gamma$ or $\theta=\beta_2$ in \eqref{eq:genlp} and \eqref{eq:genlp1} and truncating the sum in \eqref{eq:genlp} at $k=1$, we obtain the first-order LP on $\gamma$ and on $\beta_2$, respectively. The first-order LP on $\gamma$ is written as
\begin{equation}\label{eq:lpgamma}
    A(t,z) \approx A_{\LP}^{(\gamma)}(t,z) = \Azg(t,z) \exp\left(\gamma \dfrac{\Aog \argtz}{\Azg \argtz}\right),      
\end{equation}
where $\Azg$ and $\Aog$ are given by \eqref{eq:A0gammaRP} and \eqref{eq:A1gammaRP}, respectively. The accuracy of $A_{\LP}^{(\gamma)}$ is qualitatively illustrated by the blue curve in Fig.~\ref{fig:regmod}. The first-order LP on $\beta_2$ is similarly obtained as
\begin{equation}\label{eq:lpbeta2}
    A(t,z) \approx A_{\LP}^{(\beta_2)}(t,z) = \Azb(t,z) \exp\left(\beta_2 \dfrac{\Aob \argtz}{\Azb \argtz}\right),      
\end{equation}
where $\Azb$ and $\Aob$ are given by \eqref{eq:A0b2RP} and \eqref{eq:A1b2RP}, respectively.

\subsection{Frequency Logarithmic Perturbation}
\label{sc:flp}

The linearity of \eqref{eq:genrp} with respect to the functions $A_k^{(\theta)}$ suggests another approach to obtain a different LP solution. The new approach consists of performing the same steps as in \eqref{eq:genrp}--\eqref{eq:genlp1} in the frequency domain, which we refer to as FLP. To obtain the FLP solution, we first express \eqref{eq:genrp} in the frequency domain, i.e.,
\begin{equation}\label{eq:frpgen}
         \Tilde{A}(\omega,z) = \sum_{k=0}^{\infty} \theta^k \Tilde{A}_{k}^{(\theta)}(\omega,z),
\end{equation}
where $\Tilde{A}$ represents the Fourier transform\footnote{We define the Fourier transform of a function $A(\cdot,z)$ as $\Tilde{A}(\omega,z) \triangleq \int_{-\infty}^\infty \! A(t,z)\mathrm{e}^{+j\omega t} \text{d}t$, which depends on the angular frequency $\omega$ and is evaluated at distance $z$. The inverse Fourier transform of $\Tilde{A}(\cdot,z)$ is $A(t,z) = [1/(2\pi)] \int_{-\infty}^\infty \! \Tilde{A}(\omega,z)\mathrm{e}^{-j\omega t} \text{d}\omega$.} of $A$ and $\omega$ is the angular frequency. Analogous to \eqref{eq:genlp}, now $\TA$ is expressed in its FLP version as
\begin{equation}\label{eq:flpgen}
        \TA \argwz  = \TA_0^{(\theta)} \argwz \exp\left(\sum_{k=1}^\infty \theta^k \Tzeta_k^{(\theta)} \argwz \right),
\end{equation}
where the function $\Tzeta_k^{(\theta)}$ is the $k$-th FLP function and $\TA_0^{(\theta)}$ is the Fourier transform of the 0-th order RP function. 

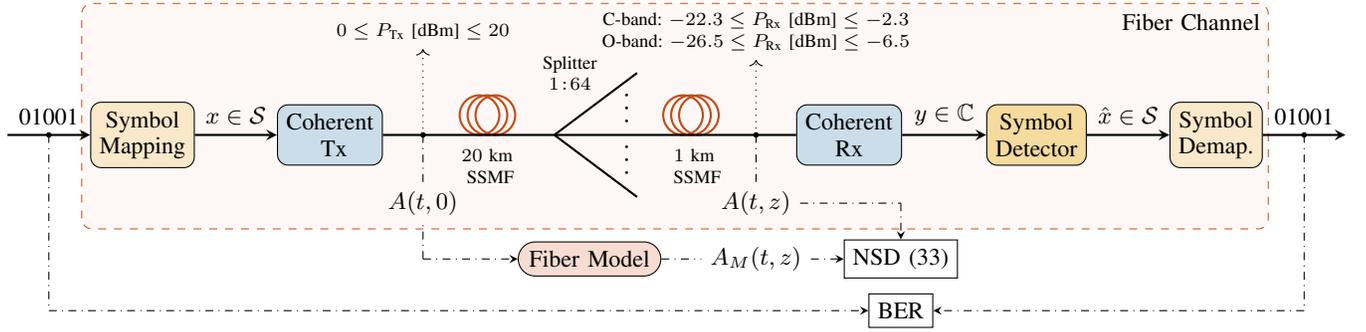
\begin{figure*}[t!]
\centering
\scalebox{1}{\setlength{\hsep}{1cm}

\tikzstyle{Cir} = [draw, circle, 
minimum size=1.5em]

\begin{tikzpicture}
\definecolor{mycolor1}{rgb}{0.00000,0.44706,0.74118}%
\definecolor{mycolor2}{rgb}{0.00000,0.49804,0.00000}%
\definecolor{mycolor3}{rgb}{0.92900,0.69400,0.12500}%
\definecolor{mycolor4}{rgb}{0.85098,0.32549,0.09804}%
\definecolor{mycolor5}{rgb}{0.39,0.19,0.0625};

\def\sca{2.75}

\def\df{0.41*\sca}
\def\dspl{0.675*\sca}
\def\dasp{0.3*\sca}
\def\dbsp{0.4*\sca}
\def\drsp{0.4*\sca}
\def\sdoUa{0.25*\sca}
\def\sdoUr{0.35*\sca}
\def\sdoDa{0.25*\sca}
\def\sdoDr{0.35*\sca}
\def\dsfibR{0.475*\sca}
\def\drxR{0.95*\sca}
\def\txB{0.15*\sca}
\def\txR{0.35*\sca}
\def\txBS{0.15*\sca}
\def\txRS{0.2*\sca}
\def\txSplB{-0.3*\sca}
\def\txSplR{0.075*\sca}
\def\ptxR{0.45*\sca}
\def\ptxA{0.5*\sca}
\def\poby{0.2*\sca}
\def\Cwd{0.03}

\def\Azd{0.2*\sca}
\def\Aend{0.25*\sca}
\def\nsdB{0.6*\sca}
\def\nsdX{0.7*\sca}
\def\berB{0.85*\sca}
\def\symbI{0.65*\sca}
\def\demP{0.65*\sca}
\def\symbO{0.625*\sca}

\def\mqam{0.4*\sca}
\def\mqamout{0.4*\sca}

\coordinate (CO) at (0,0);
\coordinate [right=\df of CO] (Fiber); 
\coordinate [right=\dspl of Fiber] (Spl);
\coordinate [above right=\dasp and \drsp of Spl] (ASpl);
\coordinate [below right=\dasp and \drsp of Spl] (BSpl);

\coordinate [above right=\sdoUa and \sdoUr of Spl] (DAb);
\coordinate [below right=\sdoDa and \sdoDr of Spl] (DBe);
\coordinate [right=\sdoDr of Spl] (DMi);

\coordinate [right=\dsfibR of Spl] (SFib);
\coordinate [right=\drxR of SFib] (Rx);

\coordinate [below right=\txB and \txR of Fiber] (TxtF1); 
\coordinate [below right=\txBS and \txRS of SFib] (TxtF2); 

\coordinate [below right=\txSplB and \txSplR of Spl] (TxtSpl);

\coordinate [above right=\ptxA and \ptxR of CO] (Ptx);
\coordinate [above left=\ptxA and \ptxR of Rx] (Prx);

\coordinate [above=\poby of Prx] (Por);

\draw[dashed,rounded corners, color=mycolor4,fill=mycolor4!10!white,fill opacity=0.4] (-3.3,-1.25) rectangle (12.475,1.75);


\node [Cir, align=center, xshift=2.45em, yshift=0.75 em,mycolor4!90!black,thick] (FiberC1) at (Fiber) {};
\node [Cir, align=center, xshift=2.75em, yshift=0.75 em,mycolor4!90!black,thick] (FiberC2) at (Fiber) {};
\node [Cir, align=center, xshift=3.05em, yshift=0.75 em,mycolor4!90!black,thick] (FiberC3) at (Fiber) {};

\node[align=center] (NCO) [rectangle,draw,rounded corners,fill=mycolor1,fill opacity=0.2,text opacity = 1] at (CO) {\small{Coherent} \\[-0.2em] \small{ Tx}};

\draw [thick,draw] (Spl)--(ASpl);
\draw [thick,draw] (Spl)--(BSpl);

\foreach \i in {1,2,3}
{
\def\auxvar{\i*0.25}
\coordinate (auxAb) at ($(DAb)!\auxvar!(DMi)$);
\node at (auxAb) {$\cdot$};
\coordinate (auxBe) at ($(DBe)!\auxvar!(DMi)$);
\node at (auxBe) {$\cdot$};
}

\node [Cir, align=center, xshift=1.15em, yshift=0.75 em,mycolor4!90!black,thick] (SFibC1) at (SFib) {};
\node [Cir, align=center, xshift=1.45em, yshift=0.75 em,mycolor4!90!black,thick] (SFibC2) at (SFib) {};
\node [Cir, align=center, xshift=1.75em, yshift=0.75 em,mycolor4!90!black,thick] (SFibC3) at (SFib) {};

\node[align=center] (NRx) [rectangle,draw,rounded corners,fill=mycolor1,fill opacity=0.2,text opacity = 1] at (Rx) {\small{Coherent}\\[-0.2em]\small{Rx}};

\draw[thick] (NCO.east)--(NRx.west);

\node [align=center]  at (TxtF1) {\scriptsize{$20$ km} \\ [-0.85ex]\scriptsize{ SSMF}};
\node [align=center] at (TxtF2) {\scriptsize{$1$ km} \\  [-0.85ex]\scriptsize{ SSMF}};

\node[align=center] at (TxtSpl) {\scriptsize{Splitter} \\ [-0.85ex] \scriptsize{$1 \! : \! 64$}};

\node (Ptxtxt) at (Ptx) {\scriptsize{$0 \le P_{\text{Tx}} \text{ [dBm]} \le 20$}};
\draw [dotted,->] (Ptxtxt |- CO) -- (Ptxtxt.south); 
\draw[fill=black] (Ptxtxt |- CO) circle (\Cwd); 

\node[align=center] (Prxtxt) at (Prx) {\scriptsize{C-band: $-22.3 \le P_{\text{Rx}} \text{ [dBm]} \le -2.3$}\\[-0.85ex]\scriptsize{O-band: $-26.5 \le P_{\text{Rx}} \text{ [dBm]} \le -6.5$}};


\draw [dotted,->] (Prxtxt |- CO) -- (Prxtxt.south); 
\draw[fill=black] (Prxtxt |- CO) circle (\Cwd); 

\coordinate (nsdL) at ($(Ptxtxt |- CO)+(0,-\nsdB)$);
\coordinate (nsdR) at (Prxtxt |- CO |- nsdL);


\node (a0txt) at ($(Ptxtxt |- CO)!0.55!(nsdL)$) {\small{$A(t,0)$}}; 

\node (aztxt) at ($(Prxtxt |- CO)!0.55!(nsdR)$) {\small{$A(t,z)$}}; 

\draw[dash dot] (Ptxtxt |- CO)--(a0txt);
\draw[dash dot] (Prxtxt |- CO)--(aztxt);
\draw[dash dot] (nsdL)--(a0txt);

\node[rounded rectangle,draw,fill=mycolor4,fill opacity=0.2,text opacity = 1] (FibMod) at ($(nsdL)!0.5!(nsdR)$) {\small{Fiber Model}};

\draw[dash dot,>=stealth,->] (nsdL) -- (FibMod.west);

\node (aMtxt) at (nsdR) {\small{$A_M(t,z)$}}; 
\draw[dash dot] (FibMod.east) -- (aMtxt);

\node[rectangle,draw,fill=white,fill opacity=0.2,text opacity = 1] (NSDtxt) at ($(nsdR)+(\nsdX,0)$) {\small{NSD (33)}};

\draw[dash dot,>=stealth,->] (aMtxt)--(NSDtxt.west);
\draw[dash dot] (aztxt.east) -- (NSDtxt|-aztxt);
\draw[dash dot,>=stealth,->] (NSDtxt|-aztxt) -- (NSDtxt.north);



\coordinate (ini) at ($(NCO.west)+(-\symbI,0)$);

\node[align=center] (NCCO) [rectangle,draw,rounded corners,fill=mycolor3,fill opacity=0.2,text opacity = 1] at (ini) {\small{Symbol} \\[-0.2em] \small{Mapping}};

\draw[thick,->,>=stealth] (NCCO.east)-- node[above] {\small{$x \in \mathcal{S}$}} (NCO);

\coordinate (iniini) at ($(NCCO.west)+(-\mqam,0)$);

\draw[thick,->,>=stealth] (iniini)-- node[above] {\small{01001}} (NCCO);

\node[align=center] (Dem) [rectangle,draw,rounded corners,fill=mycolor3,fill opacity=0.4,text opacity = 1] at ($(NRx.east)+(\demP,0)$) {\small{Symbol} \\[-0.2em] \small{Detector}};

\draw[thick,->,>=stealth] (NRx.east)-- node[above] {\small{$y \in \mathbb{C}$}} (Dem.west);

\coordinate (out) at ($(Dem.east)+(\symbO,0)$);

\node[align=center] (OUCO) [rectangle,draw,rounded corners,fill=mycolor3,fill opacity=0.2,text opacity = 1] at (out) {\small{Symbol} \\[-0.2em] \small{Demap.}};

\coordinate (outout) at ($(OUCO.east)+(\mqamout,0)$);

\draw[thick,->,>=stealth] (OUCO.east)-- node[above] {\small{01001}} (outout);

\draw[thick,->,>=stealth] (Dem.east)-- node[above] {\small{$\hat{x} \in \mathcal{S}$}} (OUCO.west);

\coordinate (midX) at ($(iniini)!0.5!(NCCO.west)$);
\coordinate (berL) at ($(midX)+(0,-\berB)$);
\coordinate (midY) at ($(outout)!0.5!(OUCO.east)$);
\coordinate (berR) at ($(midY)+(0,-\berB)$);

\draw[fill=black] (midX) circle (\Cwd); 
\draw[fill=black] (midY) circle (\Cwd); 

\draw[dash dot] (midX)--(berL);
\draw[dash dot] (midY)--(berR);
\node[rectangle,draw,fill=white,fill opacity=0.2,text opacity = 1] (BERtxt) at ($(NSDtxt)+(0,\nsdB-\berB)$) {\small{BER}};

\draw[dash dot,->,>=stealth] (berL)--(BERtxt);
\draw[dash dot,->,>=stealth] (berR)--(BERtxt);


\node at (11.45,1.5) {\small{Fiber Channel}};











\end{tikzpicture}}
\caption{PON system setup used for the simulations in this paper. This system presents low accumulated dispersion and operates in the highly nonlinear regime for the used range of input powers. The NSD is calculated using the fiber output $A$, obtained via SSFM, and a fiber model output $A_M$. The NSD exact formula given by \eqref{eq:nsd}. The BER is estimated using the bits corresponding to the input symbols $x$ and the bits corresponding to the estimated input symbols $\hat{x}$, where $x,\hat{x} \in \mathcal{S}$ and $\mathcal{S}$ is the set of constellation points. Demap.: demapping.}
\label{fig:ftthsyst}
\end{figure*}

In complete analogy with the procedure used to obtain
\eqref{eq:genlp1}, the first-order FLP function is 
\begin{equation}\label{eq:flpfun1}
    \Tzeta_1^{(\theta)} \argwz = \dfrac{\TA_1^{(\theta)} \argwz}{\TA_0^{(\theta)} \argwz},
\end{equation}
which is used to obtain the first-order FLP on $\gamma$ and on $\beta_2$. The first-order FLP on $\gamma$ is
\begin{equation}\label{eq:flpgamma}
      \TA(\omega,z) \approx \TA_{\FLP}^{(\gamma)}\argwz = \TA_0^{(\gamma)} \argwz \exp\left(\gamma \dfrac{\TA_1^{(\gamma)} \argwz}{\TA_0^{(\gamma)} \argwz}\right),  
\end{equation}
where $\TA_0^{(\gamma)}$ and $\TA_1^{(\gamma)}$ are the Fourier transforms of \eqref{eq:A0gammaRP} and \eqref{eq:A1gammaRP}, respectively. The first-order FLP on $\beta_2$ is given by
\begin{equation}\label{eq:flpbeta2}
    \TA(\omega,z) \approx \TA_{\FLP}^{(\beta_2)}\argwz = \TA_0^{(\beta_2)}\argwz \exp\left(\beta_2 \dfrac{\TA_1^{(\beta_2)} \argwz}{\TA_0^{(\beta_2)} \argwz}\right),      
\end{equation}
where $\TA_0^{(\beta_2)}$ and $\TA_1^{(\beta_2)}$ are the Fourier transforms of \eqref{eq:A0b2RP} and \eqref{eq:A1b2RP}, respectively. The qualitative behaviour of the accuracy of $\TA_{\FLP}^{(\beta_2)}$ is illustrated as the green dashed curve in Fig.~\ref{fig:regmod}.

The functions $\Tzeta_k^{(\theta)}$ in \eqref{eq:flpgen} differ from the Fourier transform of $\psi_k^{(\theta)}$ in \eqref{eq:genlp}, since the exponential of the LP method was applied in the frequency domain. The expressions in \eqref{eq:genlp1} and \eqref{eq:flpfun1} are the simplest example of this fact, since they do not form, in general, a Fourier transform pair. Therefore, we expect that these LP-based models result in a different accuracy for each perturbation coefficient ($\gamma$ or $\beta_2$). As it will be seen later in Sec.~\ref{sc:setup}, $A_{\LP}^{(\gamma)}$ is more accurate than $A_{\FLP}^{(\gamma)}$, while $A_{\FLP}^{(\beta_2)}$ is more accurate than $A_{\LP}^{(\beta_2)}$. We believe that the difference between $\beta_2$ and $\gamma$ when comparing LP and FLP could be explained by the solution of \eqref{eq:normnlse} for only the chromatic dispersion effect or only the Kerr effect \cite{AGRAWAL201327}. The solution for the chromatic dispersion effect only is an exponential in the frequency domain, which resembles the FLP approach. Similarly, the solution for the Kerr effect only is an exponential in the time domain, which resembles the LP approach.

Together with the models in the previous sections, we obtained six perturbation models: two RPs in \eqref{eq:rpg1} and \eqref{eq:rpb2_1}; two LPs in \eqref{eq:lpgamma} and \eqref{eq:lpbeta2}; and two FLPs in \eqref{eq:flpgamma} and \eqref{eq:flpbeta2}. Table~\ref{tab:pfunc} summarizes these six first-order perturbation methods. As shown in Table~\ref{tab:pfunc}, LP and FLP can be obtained using the RP terms.


\section{Simulation Setup and Results} \label{sc:setup}


The model presented in this work is validated in a PON transmission scenario where the accumulated dispersion is expected to be low. 
Fig.~\ref{fig:ftthsyst} shows the coherent PON system setup under consideration. The fiber parameters are given in Table~\ref{tab:param}. As depicted in the figure, we consider a standard single mode fiber (SSMF) of $20$ km, followed by a splitter of ratio $1\!:\!64$ and a final fiber segment of $1$ km. With this split ratio, the total link loss is $22.3$ dB for the C-band and $26.5$ dB for the O-band. The fiber input power $P_{\text{Tx}}$ varies from $0$ to $20$ dBm, which leads to a received power $P_{\text{Rx}}$ between $-22.3$ and $-2.3$ dBm in the C-band and between $-26.5$ and $-6.5$ dBm in the O-band. The range of powers was chosen to cover and go beyond launch powers for typical PON systems according to \cite{ITU2016,ITU2018}. All the results were obtained using randomly generated bits which were mapped into symbols drawn from a quadrature phase shift keying (QPSK) constellation $\mathcal{S} = (\pm1 \pm j)/\sqrt{2}$. The coherent transmitter applies pulse shaping and scales the waveform such that the average transmitted power is $P_\text{Tx}$. The coherent receiver undo the waveform scaling, and then applies matched filtering and sampling, without chromatic-dispersion compensation. The symbol rate of the transmitted signal is $10$ Gbaud for both C- and O-band systems. The considered pulse shape is a root-raised cosine (RRC), with a roll-off factor of $0.1$. We consider a noiseless scenario since for the considered bandwidth and received powers, nonlinear distortions dominate over the shot noise \cite{Lavery:15,Lavery2015comp}.

In the considered system setup, we want to evaluate the impact of nonlinearities and dispersion on the models. For evaluating the impact of the nonlinearities, the power was varied as specified before. The effect of the dispersion in the models is evaluated by comparing the C-band and the O-band scenarios, which have different $\beta_2$ values. We consider the effect of the TOD in the O-band since in that regime values of $\beta_2$ are low. Therefore, the equation used for the SSFM simulation is slightly different from \eqref{eq:normnlse}, i.e., it considers an additional term accounting for TOD \cite[Eq.\ (3.3.1)]{AGRAWAL201327}, resulting in 
 \begin{align}\label{eq:nlse3ord}
   \dfrac{\partial A(t,z)}{\partial z} = &-\dfrac{j\beta_2}{2}\dfrac{\partial^2 A(t,z)}{\partial t^2 } + \dfrac{\beta_3}{6}\dfrac{\partial^3 A(t,z)}{\partial t^3}  \notag \\ &+ j\gamma \mathrm{e}^{-\alpha z}|A(t,z)|^2 A(t,z).
 \end{align}


We evaluate the model accuracy on two levels, comparing either the channel output waveforms or detected symbols. For the former, no receiver is considered. The accuracy is quantified using the normalized squared deviation (NSD) metric \cite{Vannucci2002,Oliari2020}

\begin{equation}\label{eq:nsd}
    \text{NSD} \triangleq \dfrac{\int_{-\infty}^{\infty} \! |A_M(t,z)-A(t,z)|^2 \text{d}t}{\int_{-\infty}^{\infty} |A(t,z)|^2\text{d}t},
\end{equation}
where $A_M$ is a model output (i.e., $A_{\LP}^{(\gamma)}$ or $A_{\FLP}^{(\beta_2)}$, for example) and $A$ is the \textit{true} fiber output obtained from the SSFM. The NSD integrates the absolute error squared over the entire signal duration, and normalizes it with the energy of the signal $A$. Therefore, the lower the NSD, the more accurate is the waveform predicted by the model. The inputs for the NSD calculation are illustrated in Fig.~\eqref{fig:ftthsyst}.

\begin{table}[t!]
\centering
\caption{Fiber parameters for C- and O-band transmission}
\begin{tabular}{|c || c | c |} 
\hline
 Parameter & C-band & O-band \\ 
 \hline
 Wavelength $\lambda$ [nm]  & 1550 & 1310 \\
$\alpha$ [dB/km]  & $0.2$ &$0.4$  \\
$\beta_2$ [ps$^2$/km] & $-21.67$ & $-0.2$ \\
$\gamma$ [$1$/W/km] & $1.2$ &$1.4$ \\
$\beta_3$ [ps$^3$/km] &  $-$  & $0.0765$\\
\hline
\end{tabular}
\label{tab:param}
\end{table}

Calculating the waveforms for both LP and FLP leads to a numerical issue related to the ratio in \eqref{eq:genlp1} and \eqref{eq:flpfun1}. When the denominator in one of those two equations tends to zero, the respective model becomes inaccurate. To address this problem in the LP case, \cite{Ciaramella2005} proposed to replace $A_{\LP}^{(\theta)}(t,z)$ by $A_{\RP}^{(\theta)}(t,z)$ whenever $|A_0(t,z)|<\epsilon$, where $\epsilon>0$ is a fixed threshold. For the FLP case, we use the same technique in the frequency domain, replacing $\TA_{\FLP}^{( \theta)}\argwz $ by $\TA_{\RP}^{( \theta)} \argwz$ whenever $|\TA_0^{(\theta)} \argwz|<\epsilon$ for a certain real $\epsilon>0$. In addition, $\TA_{\FLP}^{(\theta)} \argwz $ is also replaced with $\TA_{\RP}^{(\theta)} \argwz$ whenever $|\TA_{\FLP}^{(\theta)} \argwz | > c \, |\TA_{\RP }^{(\theta)} \argwz|$ for a certain fixed real $c>1$. In order to increase the stability of the LP solution, we determined $c$ similarly to the LP case. For the results presented in this paper $\epsilon$ and $c$ are heuristically fixed to $c=1.15$ and $\epsilon=\text{max}_t\{|A_0^{(\gamma)}(t,z)|\}/10^5$ for LP on $\gamma$ and $c=1.1$ and $\epsilon=\text{max}_\omega\{|\TA_0^{(\beta_2)}(\omega,z)\}/10^6$ for FLP on $\beta_2$.

For the symbol-level evaluation, decision regions are optimized according to each model as described in Sec.~\ref{sc:decreg}. In that section, the accuracy is quantified in terms of BER (illustrated in Fig.~\ref{fig:ftthsyst}), while in Sec.~\ref{sc:achivrat}, the accuracy is quantified in terms of achievable information rate (AIR).






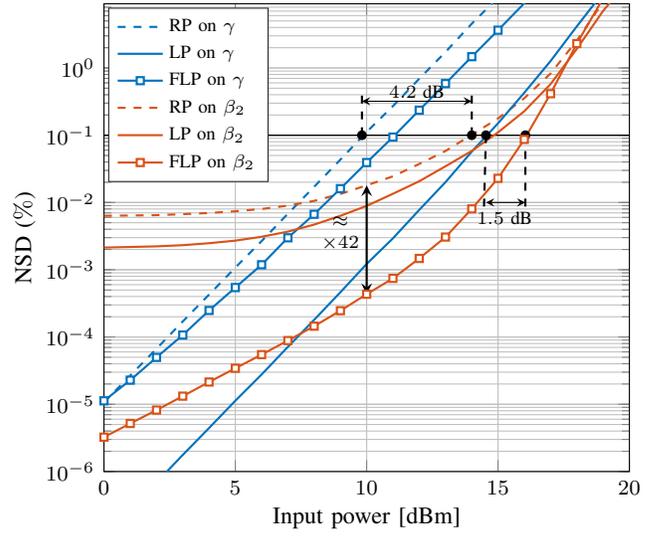
\begin{figure}[!t]
\centering
    
\def\slo{\small}
\def\to{\footnotesize}

\def\xsiz{1.4}
\def\linW{0.8}

\definecolor{mycolor1}{rgb}{0.00000,0.44700,0.74100}%
\definecolor{mycolor2}{rgb}{0.00000,0.49804,0.00000}%
\definecolor{mycolor3}{rgb}{0.92900,0.69400,0.12500}%
\definecolor{mycolor4}{rgb}{0.85098,0.32549,0.09804}%
\definecolor{mycolor5}{rgb}{0.49400,0.18400,0.55600}%
\definecolor{mycolor6}{rgb}{0.0,0.1,0.08}%
\definecolor{mycolor7}{rgb}{0.00,0.6400,0.512}%

\begin{tikzpicture}

\begin{axis}[%
width=2.75in,
height=2.45in,
at={(0in,0in)},
scale only axis,
xmin=0,
xmax=20,
xlabel style={font=\color{white!15!black},font=\slo},
xlabel={Input power [dBm]},
ymode=log,
ymin=1e-6,
ymax=9,
yminorticks=true,
ylabel style={font=\color{white!15!black},font=\slo},
ylabel={NSD ($\%$)},
axis background/.style={fill=white},
xlabel shift = -0.03in,
ylabel shift = -0.04in,
xmajorgrids,
ymajorgrids,
yminorgrids,
ticklabel style = {font=\to},
 legend columns=2, 
legend style={at={(0.42,0.03)}, anchor=south west, legend cell align=left, align=left, draw=white!15!black,font=\scriptsize,            /tikz/column 2/.style={
                column sep=5pt,
            }}
]








\addplot [color=black, solid, line width=0.5pt,forget plot]
table[row sep=crcr]{%
-10	0.1\\
25	0.1\\
};


\addplot [color=mycolor1,mark size=0.9pt, line width=0.8pt,  dashed, mark options={fill=white,solid,scale=1.5}]
  table[row sep=crcr]{%
0	1.06918514777054e-05\\
1	2.67002850544669e-05\\
2	6.70000095270892e-05\\
3	0.000170493840864442\\
4	0.00042667474471988\\
5	0.00107646431306764\\
6	0.00266150430688387\\
7	0.00680158360426306\\
8	0.0171030436279441\\
9	0.0433818687368914\\
10	0.111997784495301\\
11	0.270916616115128\\
12	0.690697805459424\\
13	1.7263172208747\\
14	4.55326024168876\\
15	11.1308740852893\\
16	27.4522087204434\\
17	64.444023134941\\
18	142.887145842781\\
19	288.229494681894\\
20	523.042503848167\\
};

\addplot [color=mycolor4,mark size=0.9pt, line width=0.8pt, dashed, mark options={fill=white,solid,scale=1.5}]
  table[row sep=crcr]{%
0	0.00634685039545183\\
1	0.00640704269161855\\
2	0.00648218262266783\\
3	0.00669956743003607\\
4	0.00697660848722829\\
5	0.0074073581052987\\
6	0.00801314636789469\\
7	0.00896409164657759\\
8	0.0105713569984076\\
9	0.0134020321833625\\
10	0.0179910440939485\\
11	0.0244686087187289\\
12	0.0373906654060248\\
13	0.058286046237853\\
14	0.101505813294772\\
15	0.177491529998745\\
16	0.35496158660275\\
17	0.826682328260798\\
18	2.55416458527332\\
19	9.61914678154712\\
20	29.6861119356774\\
};

\addplot [color=mycolor1,mark size=0.9pt, line width=0.8pt,   mark options={fill=white,solid,scale=1.5}]
  table[row sep=crcr]{%
0	1.10708455837968e-07\\
1	2.74781777397243e-07\\
2	6.8850181915758e-07\\
3	1.76572506036266e-06\\
4	4.43246126188175e-06\\
5	1.12738129504459e-05\\
6	2.76883145047434e-05\\
7	7.12835979854113e-05\\
8	0.000179897309135048\\
9	0.000462722670313027\\
10	0.00122040156559962\\
11	0.00293179959965093\\
12	0.00775378563785864\\
13	0.0200531127318595\\
14	0.0570112131402711\\
15	0.150263076656386\\
16	0.428244929157131\\
17	1.26731065509521\\
18	4.15698452988104\\
19	13.2125377616577\\
20	31.6044899728889\\
};

\addplot [color=mycolor4,mark size=0.9pt, line width=0.8pt,  mark=square*, mark options={fill=white,solid,scale=1.5}]
  table[row sep=crcr]{%
0	3.24411354936043e-06\\
1	5.17537098317128e-06\\
2	8.19241217561374e-06\\
3	1.32004036702838e-05\\
4	2.14118352324082e-05\\
5	3.42872064667159e-05\\
6	5.47399384991079e-05\\
7	8.85222961574623e-05\\
8	0.000144773799717005\\
9	0.00024602313367965\\
10	0.000432343844090067\\
11	0.000744208844889535\\
12	0.00147144581847609\\
13	0.00306144551128366\\
14	0.00803067783831388\\
15	0.0228440339534464\\
16	0.0866945051409156\\
17	0.414544647846444\\
18	2.30339701204965\\
19	10.8299334404294\\
20	33.4799415866188\\
};

\addplot [color=mycolor1,mark size=0.9pt, line width=0.8pt,  mark=square*, mark options={fill=white,solid,scale=1.5}]
  table[row sep=crcr]{%
0	1.12485633287492e-05\\
1	2.28219936278286e-05\\
2	4.97237953861232e-05\\
3	0.000107030753506064\\
4	0.000248011419179587\\
5	0.000543539127298167\\
6	0.00118656236174636\\
7	0.00298388004996568\\
8	0.00667704549054135\\
9	0.016053599039292\\
10	0.0392500901148419\\
11	0.0938623169487657\\
12	0.235621851688684\\
13	0.58577195614615\\
14	1.47129580995679\\
15	3.6408770082787\\
16	9.16345283965336\\
17	21.3617421693738\\
18	49.038069276756\\
19	93.6189293833627\\
20	157.274791707689\\
};

\addplot [color=mycolor4,mark size=0.9pt, line width=0.8pt]
  table[row sep=crcr]{%
0	0.00213625899248028\\
1	0.00217905997265815\\
2	0.00222776317552424\\
3	0.00231817583007248\\
4	0.00248103234454557\\
5	0.00270888107902884\\
6	0.00309232404917496\\
7	0.00368662651388598\\
8	0.00468934266451801\\
9	0.00633371042946056\\
10	0.00891031523923157\\
11	0.0134763155129678\\
12	0.0206878888211764\\
13	0.034537821989245\\
14	0.0583879969706661\\
15	0.110229007986663\\
16	0.227360767305902\\
17	0.575241893230802\\
18	1.92339213650921\\
19	7.05934005564354\\
20	20.5280399215606\\
};

\draw[thick,dashed] (9.82,0.1) -- (9.82,0.4);
\draw[thick,dashed] (14,0.1) -- (14,0.4);

\draw [<->,>=stealth] (9.82,0.32)--(14,0.32);

\node (gap101) [align=center] at (11.91,0.45) {\scriptsize{$4.2$ dB}};

\draw[thick,dashed] (14.54,0.1) -- (14.46,0.007);
\draw[thick,dashed] (16.04,0.1) -- (16.04,0.007);

\draw [<->,>=stealth] (14.54,0.01)--(16.04,0.01);

\node (gap102) [align=center] at (15.25,0.006) {\scriptsize{$1.5$ dB}};

\node [inner sep=1.3,circle,fill=black] at (9.82,0.1){};
\node [inner sep=1.3,circle,fill=black] at (14,0.1){};
\node [inner sep=1.3,circle,fill=black] at (14.54,0.1){};
\node [inner sep=1.3,circle,fill=black] at (16.04,0.1){};


\draw [<->,>=stealth, thick] (10,0.0179910440939485)--(10,0.000432343844090067);

\node (gap103) [align=center] at (9,0.0035) {\scriptsize{$\approx$} \\ [-0.85ex]  \scriptsize{$ \times 42$}};

\end{axis}

\begin{axis}[%
width=0.34 in,
height=0.35 in,
at={(0.09in,1.875in)},
scale only axis,
xmin=16,
xmax=17,
xlabel style={font=\color{white!15!black},font=\fontsize{\slo pt}{8}\sffamily},
ymin=9,
ymax=15,
yminorticks=true,
ylabel style={font=\color{white!15!black},font=\fontsize{\slo pt}{8}\sffamily},
axis background/.style={fill=white},
xlabel shift = -0.03in,
ylabel shift = -0.04in,
ticks = none,
legend style={at={(-0.2,-1.025)}, anchor=south west, legend cell align=left, align=left, draw=white!15!black,font=\scriptsize,            /tikz/column 2/.style={
                column sep=5pt,
            }}
]

\addplot[color=mycolor1,mark size=0.9pt, line width=0.8pt,  dashed, mark options={fill=white,solid,scale=1.5}]
  table[row sep=crcr]{%
15	13.6041240833658\\
};
\addlegendentry{RP on $\gamma$}

\addplot  [color=mycolor1,mark size=0.9pt, line width=0.8pt,   mark options={fill=white,solid,scale=1.5}]
  table[row sep=crcr]{%
15	13.7325917615983\\
};
\addlegendentry{LP on $\gamma$}

\addplot [color=mycolor1,mark size=0.9pt, line width=0.8pt,  mark=square*, mark options={fill=white,solid,scale=1.5}]
  table[row sep=crcr]{%
0	1.12485633287492e-05\\
};
\addlegendentry{FLP on $\gamma$}

\addplot  [color=mycolor4,mark size=0.9pt, line width=0.8pt, dashed, mark options={fill=white,solid,scale=1.5}]
  table[row sep=crcr]{%
15	13.6080629448026\\
};
\addlegendentry{RP on $\beta_2$}

\addplot [color=mycolor4,mark size=0.9pt, line width=0.8pt]
  table[row sep=crcr]{%
0	0.00213625899248028\\
};
\addlegendentry{LP on $\beta_2$}

\addplot [color=mycolor4,mark size=0.9pt, line width=0.8pt,  mark=square*, mark options={fill=white,solid,scale=1.5}]
  table[row sep=crcr]{%
15	13.7366856594518\\
};
\addlegendentry{FLP on $\beta_2$}


\end{axis}

\end{tikzpicture}
\caption{NSD for for RP, LP and FLP on $\gamma$ and on $\beta_2$ in the C-band using the system in Fig.~\ref{fig:ftthsyst}. The fiber parameters are given in Table~\ref{tab:param}.  
}
\label{fig:nsd-pvar}
\end{figure}

\subsection{Waveform Comparison}

Fig.~\ref{fig:nsd-pvar} shows the NSD for RP, LP, and FLP on $\gamma$ and on $\beta_2$. As depicted in Fig.~\ref{fig:nsd-pvar}, FLP on $\beta_2$ (solid red line with squares) is the most accurate at powers higher than $7.5$ dBm and NSD below $0.1\%$, while LP on $\gamma$ (solid blue line) is the most accurate at powers lower than $7.5$ dBm. FLP on $\gamma$ (solid blue line with squares) and LP on $\beta_2$ (solid red line) have a slightly better performance than RP on $\gamma$ (dashed blue line) and RP on $\beta_2$ (dashed red line), respectively. However, we do not consider FLP on $\gamma$ and LP on $\beta_2$ further in this paper, since their performance is surpassed by LP on $\gamma$ and FLP on $\beta_2$, respectively. RP on $\beta_2$ crosses the 0.1$\%$ NSD line at an input power approximately 4.2 dB higher than RP on $\gamma$, at 14 and 9.8 dBm, respectively. This gap in favor of RP on $\beta_2$ was expected since input powers greater than 10 dBm and small distances such as 20 km put the fiber in the highly nonlinear regime with low accumulated dispersion. As discussed in \cite{Oliari2020}, RP on $\beta_2$ is accurate on this regime, while RP on $\gamma$ loses accuracy at high powers. If we change from RP on $\gamma$ to LP on $\gamma$, the latter outperforms RP on $\beta_2$ for powers below 16 dBm. This gain in accuracy by changing from the LP on $\gamma$ to the RP on $\gamma$ was previously shown in \cite{Ciaramella2005}. In addition, LP on $\gamma$ has its performance increased due to low accumulated dispersion. By letting $\beta_2 \to 0$ in \eqref{eq:lpgamma}, LP on $\gamma$ tends to the NLPN solution, which is accurate in very low dispersion scenarios \cite{Oliari2020}.

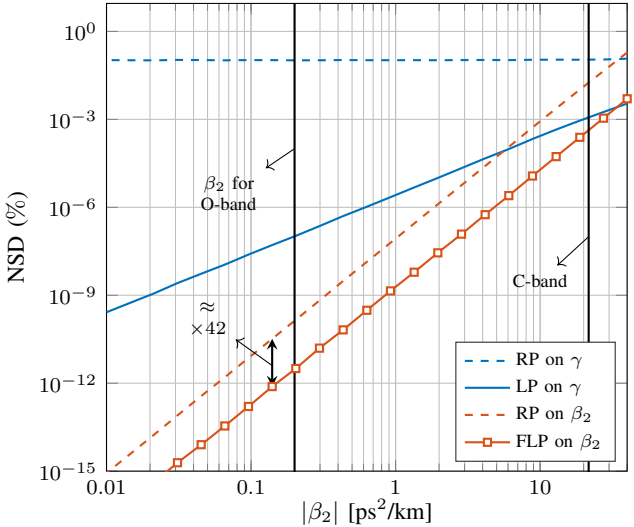
\begin{figure}[!t]
\centering
    
\def\slo{\small}
\def\to{\footnotesize}

\definecolor{mycolor1}{rgb}{0.00000,0.44700,0.74100}%
\definecolor{mycolor2}{rgb}{0.00000,0.49804,0.00000}%
\definecolor{mycolor3}{rgb}{0.92900,0.69400,0.12500}%
\definecolor{mycolor4}{rgb}{0.85098,0.32549,0.09804}%
\definecolor{mycolor5}{rgb}{0.00000,0.44706,0.74118}%
\definecolor{mycolor6}{rgb}{0.92941,0.69412,0.12549}%

\begin{tikzpicture}

\begin{axis}[%
width=2.725in,
height=2.45in,
at={(3.5in,0in)},
scale only axis,
xmode=log,
xminorticks=true,
xmajorgrids,
xminorgrids,
xmin=0.01, xmax=40, 
xticklabel=\pgfmathparse{exp(\tick)}\pgfmathprintnumber{\pgfmathresult}, 
xticklabel style={/pgf/number format/.cd,fixed}, 
xlabel style={font=\color{white!15!black},font=\slo},
xlabel={$|\beta_2|$ [ps$^2$/km]},
ymode=log,
ymin=1e-15,
ymax=9,
yminorticks=true,
ylabel style={font=\color{white!15!black},font=\slo},
ylabel={NSD ($\%$)},
axis background/.style={fill=white},
xlabel shift = -0.07in,
ylabel shift = -0.04in,
xmajorgrids,
ymajorgrids,
yminorgrids,
ticklabel style = {font=\to},
legend style={at={(0.67,0.02)}, anchor=south west, legend cell align=left, align=left, draw=white!15!black,font=\scriptsize,            /tikz/column 2/.style={
                column sep=5pt,
            }}
]












\addplot[color=mycolor1,mark size=0.9pt, line width=0.8pt,  dashed, mark options={fill=white,solid,scale=1.5}]
table[row sep=crcr]{%
40	0.117241861114121\\
27.43658104874	0.110014864557904\\
18.8191494911019	0.108022018260705\\
12.9083280070242	0.108610614289014\\
8.85400968921101	0.107635862246447\\
6.07309386110917	0.104816320041146\\
4.16562329842317	0.106188072055805\\
2.85726153114266	0.105114283286141\\
1.95983718941606	0.103973742481627\\
1.34428079724371	0.104515300441397\\
0.922061726146046	0.104192498231485\\
0.632455532033676	0.104165445767978\\
0.433810436609147	0.105614390287044\\
0.297556880095403	0.103452693045808\\
0.204098586433693	0.102845209971932\\
0.139994185215532	0.104615446300436\\
0.0960240452254562	0.104961566543014\\
0.0658642874864024	0.103053531767958\\
0.0451772715459547	0.105343954200993\\
0.030987746808288	0.106627852060346\\
0.0212549456705857	0.102671659302817\\
0.0145790759894397	0.103663563814778\\
0.01	0.105071735155488\\
};
\addlegendentry{RP on $\gamma$}

\addplot  [color=mycolor1,mark size=0.9pt, line width=0.8pt,   mark options={fill=white,solid,scale=1.5}]
  table[row sep=crcr]{%
40	0.00345552731694695\\
27.43658104874	0.00177926608711166\\
18.8191494911019	0.000904183580815635\\
12.9083280070242	0.000448457674464001\\
8.85400968921101	0.000213594504041689\\
6.07309386110917	9.68994882085907e-05\\
4.16562329842317	4.67370005995315e-05\\
2.85726153114266	2.16062797224311e-05\\
1.95983718941606	1.00327026741096e-05\\
1.34428079724371	4.75068150676888e-06\\
0.922061726146046	2.21785358902142e-06\\
0.632455532033676	1.04401788161776e-06\\
0.433810436609147	5.0208399388583e-07\\
0.297556880095403	2.28565666187008e-07\\
0.204098586433693	1.06201020025043e-07\\
0.139994185215532	5.14369431571203e-08\\
0.0960240452254562	2.44135276075832e-08\\
0.0658642874864024	1.10833299167577e-08\\
0.0451772715459547	5.40136033317108e-09\\
0.030987746808288	2.59858008391646e-09\\
0.0212549456705857	1.14913398106862e-09\\
0.0145790759894397	5.51295101277121e-10\\
0.01	2.63237063631144e-10\\
};
\addlegendentry{LP on $\gamma$}

\addplot  [color=mycolor4,mark size=0.9pt, line width=0.8pt, dashed, mark options={fill=white,solid,scale=1.5}]
  table[row sep=crcr]{%
40	0.190497669895068\\
27.43658104874	0.0443552103411358\\
18.8191494911019	0.010378886316081\\
12.9083280070242	0.00234921825949286\\
8.85400968921101	0.000528790990588975\\
6.07309386110917	0.000114501688329791\\
4.16562329842317	2.55866808189133e-05\\
2.85726153114266	5.65407837939597e-06\\
1.95983718941606	1.26687479149731e-06\\
1.34428079724371	2.79457663642473e-07\\
0.922061726146046	6.15509135736492e-08\\
0.632455532033676	1.37557901893541e-08\\
0.433810436609147	3.07267153058651e-09\\
0.297556880095403	6.62855884374693e-10\\
0.204098586433693	1.457991860833e-10\\
0.139994185215532	3.26656809100153e-11\\
0.0960240452254562	7.27443655069238e-12\\
0.0658642874864024	1.59625882954609e-12\\
0.0451772715459547	3.63564419705212e-13\\
0.030987746808288	7.99858739262705e-14\\
0.0212549456705857	1.72411852432069e-14\\
0.0145790759894397	3.84504050346555e-15\\
0.01	8.6662477748825e-16\\
};
\addlegendentry{RP on $\beta_2$}

\addplot [color=mycolor4,mark size=0.9pt, line width=0.8pt,  mark=square*, mark options={fill=white,solid,scale=1.5}]
  table[row sep=crcr]{%
40	0.00503744297016073\\
27.43658104874	0.00109507207962645\\
18.8191494911019	0.000243507661472864\\
12.9083280070242	5.35109838796171e-05\\
8.85400968921101	1.17005394529241e-05\\
6.07309386110917	2.50230579650999e-06\\
4.16562329842317	5.60927238549978e-07\\
2.85726153114266	1.21606365918952e-07\\
1.95983718941606	2.79952042209115e-08\\
1.34428079724371	6.11615361099643e-09\\
0.922061726146046	1.40502403776118e-09\\
0.632455532033676	3.07279124367479e-10\\
0.433810436609147	6.58695016360199e-11\\
0.297556880095403	1.56914700754172e-11\\
0.204098586433693	3.16723228897366e-12\\
0.139994185215532	7.67112153161872e-13\\
0.0960240452254562	1.60824248151056e-13\\
0.0658642874864024	3.48343681912392e-14\\
0.0451772715459547	7.98052868444265e-15\\
0.030987746808288	1.93747495042606e-15\\
0.0212549456705857	4.14014212271131e-16\\
0.0145790759894397	9.1793591493584e-17\\
0.01	2.11217044420553e-17\\
};
\addlegendentry{FLP on $\beta_2$}

\draw[thick] (21.67,10) -- (21.67,1e-16);
\draw[thick]  (0.2,10) -- (0.2,1e-16);

\node (cb) at (10,3e-9) {\scriptsize{C-band}};
\draw [->] (21.67,1e-7) -- (cb);

\node[align=center] (ob) at (0.07,3e-6) {\scriptsize{$\beta_2$ for}\\ [-0.85ex] \scriptsize{O-band}};
\draw [->] (0.2,1e-4) -- (ob);

\draw [<->,>=stealth,thick] (0.14,7.67112153161872e-13)--(0.14,3.26656809100153e-11);

\node (gap42) [align=center] at (0.05,1.5e-10) {\scriptsize{$\approx$} \\ [-0.85ex]  \scriptsize{$ \times 42$}};

\draw [->] (0.14,4e-12) -- (gap42);

\end{axis}

\end{tikzpicture}
\caption{NSD versus $|\beta_2|$ (negative $\beta_2$) for four models at an input power of 10 dBm. The system is represented in Fig.~\ref{fig:ftthsyst}. All considered models except RP on $\gamma$ get higher accuracy as $|\beta_2|$ decreases. 
}
\label{fig:nsd-vs-b2}
\end{figure}

The dependence of the models on $|\beta_2|$ can be seen in Fig.~\ref{fig:nsd-vs-b2}, where four models are compared at a fixed power of 10 dBm for different values of $|\beta_2|$ and the other parameters for C-band transmission (with no TOD). All the simulated $\beta_2$ values were negative. Among the four models, RP on $\gamma$ is the only one that is virtually invariant to changes in $\beta_2$ for the displayed values. Nevertheless, RP on $\gamma$ has worse accuracy than LP on $\gamma$ for all displayed values of $\beta_2$. When increasing $|\beta_2|$, LP on $\gamma$ increases its NSD at a rate of approximately $10^2$ per decade. Although LP on $\gamma$ outperforms RP on $\beta_2$ at 10 dBm for the C-band, the latter has an increasing rate of approximately $10^4$ per decade, and surpasses the accuracy of LP on $\gamma$ for $\beta_2$ values lower than $-6$ ps$^2$/km. If we now also consider FLP on $\beta_2$, we can gain approximately 42 times in NSD accuracy with respect to RP on $\beta_2$ at 10 dBm. This gap can also be seen in Fig.~\ref{fig:nsd-pvar} and remains approximately constant for different values of $\beta_2$, since both RP and FLP on $\beta_2$ have the same increasing rate of approximately $10^4$ per decade. The higher increasing rate for RP and FLP on $\beta_2$ when compared with LP on $\gamma$ shows that the two former models converge to the true solution of \eqref{eq:normnlse} faster than the latter model when decreasing the accumulated dispersion. For $\beta_2=-21.67$ (C-band), FLP on $\beta_2$ already outperforms LP on $\gamma$. In addition, we see back in Fig.~\ref{fig:nsd-pvar} that FLP on $\beta_2$ crosses the line for an NSD of 0.1$\%$ at an input power 1.5 dB higher than LP on $\gamma$.


\begin{figure}[!t]
\centering
\input{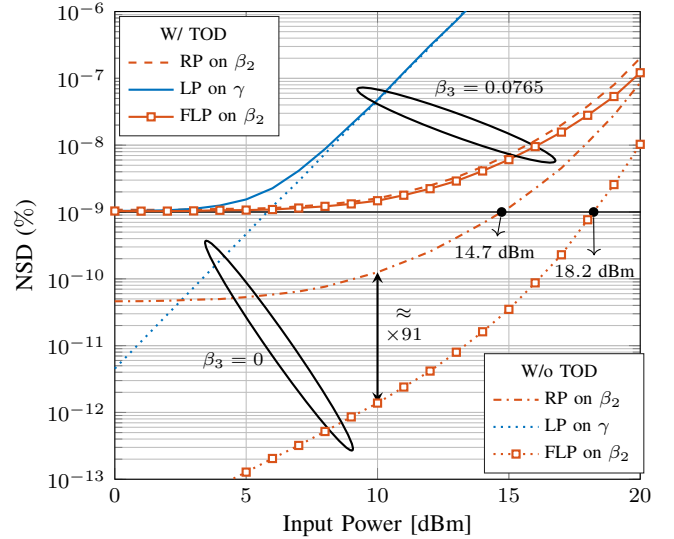}
\caption{NSD for the four models in the O-band for the system in Fig.~\ref{fig:ftthsyst}. Results without TOD are also shown. 
}
\label{fig:nsd-pvarOb}
\end{figure}

The results for LP on $\gamma$, RP on $\beta_2$, and FLP on $\beta_2$ using the O-band parameters in Table~\ref{tab:param} are shown in Fig.~\ref{fig:nsd-pvarOb}, where the NSD is displayed as a function of the input power. First, we consider a system without TOD (dotted and dashed dotted lines). As shown in Fig.~\ref{fig:nsd-pvarOb}, the NSD for LP on $\gamma$ and FLP on $\beta_2$ significantly decay when reducing the input power in the absence of TOD. On the other hand, RP on $\beta_2$ converges to an NSD of approximately $4.6\cdot 10^{-11} \%$ for powers lower than 2 dBm. This convergence to a non-zero NSD value reflects the mismatch between \eqref{eq:genrpb2} and \eqref{eq:rpb2_1} in the absence of nonlinearities. FLP on $\beta_2$ outperforms LP on $\gamma$ for all the displayed input powers, while RP on $\beta_2$ outperforms LP on $\gamma$ for input powers higher than 2.5 dBm. At 10 dBm, the difference in NSD between RP and FLP on $\beta_2$ is approximately 91 times, which is different from the 42 factor for the C-band results at the same input power (see Fig.~\ref{fig:nsd-vs-b2}). This discrepancy is due to the new set of $\gamma$ and $\alpha$ values, which boost the difference between the two models at that input power.

The results for the O-band in Fig.~\ref{fig:nsd-pvarOb} shows that when TOD is considered, the three models now converge to a constant NSD of $10^{-9}$ for input powers lower than 3 dBm. This behavior can be explained by the absence of TOD in the model derivations. The error introduced by not accounting for TOD becomes approximately constant when input powers are lower than 3 dBm, and dominates the error introduced by incorrectly modeling the other fiber effects. 
From 0 to 20 dBm, the performance of RP and FLP on $\beta_2$ in the system with TOD is worse than without TOD. This behavior is expected, since for the system without TOD, RP and FLP on $\beta_2$ cross the constant NSD of $10^{-9}$ (TOD error floor for low powers) only at 14.7 dBm and 18.2 dBm, respectively. For LP on $\gamma$, the NSD for the system without TOD already crosses the $10^{-9}$ line at 6 dBm. Therefore, results for the systems with and without TOD converge for powers greater than 10 dBm.

\subsection{Decision Region Optimization}
\label{sc:decreg}

As discussed in \cite{Oliari2020}, comparing models in discrete time can lead to slightly different conclusions than on a waveform level. For this reason, this section compares LP on $\gamma$, FLP on $\beta_2$, and SSFM results at the symbol level, measured by BER. The results are shown for the C-band system with parameters given in Table~\ref{tab:param}. To obtain the symbols, the output waveform from these three models is filtered by a matched filter and sampled as done for the signal $A(\cdot,z)$ in Fig.~\ref{fig:ftthsyst}. The resulting complex samples are used to optimize decision regions for each model, originating a symbol detector. Finally, SSFM simulations are performed to validate the accuracy of each symbol detector when receiving the \textit{true} (SSFM) output waveform. These results are also compared with the PW detector in \cite{karim2019}.


\begin{figure*}[!t]
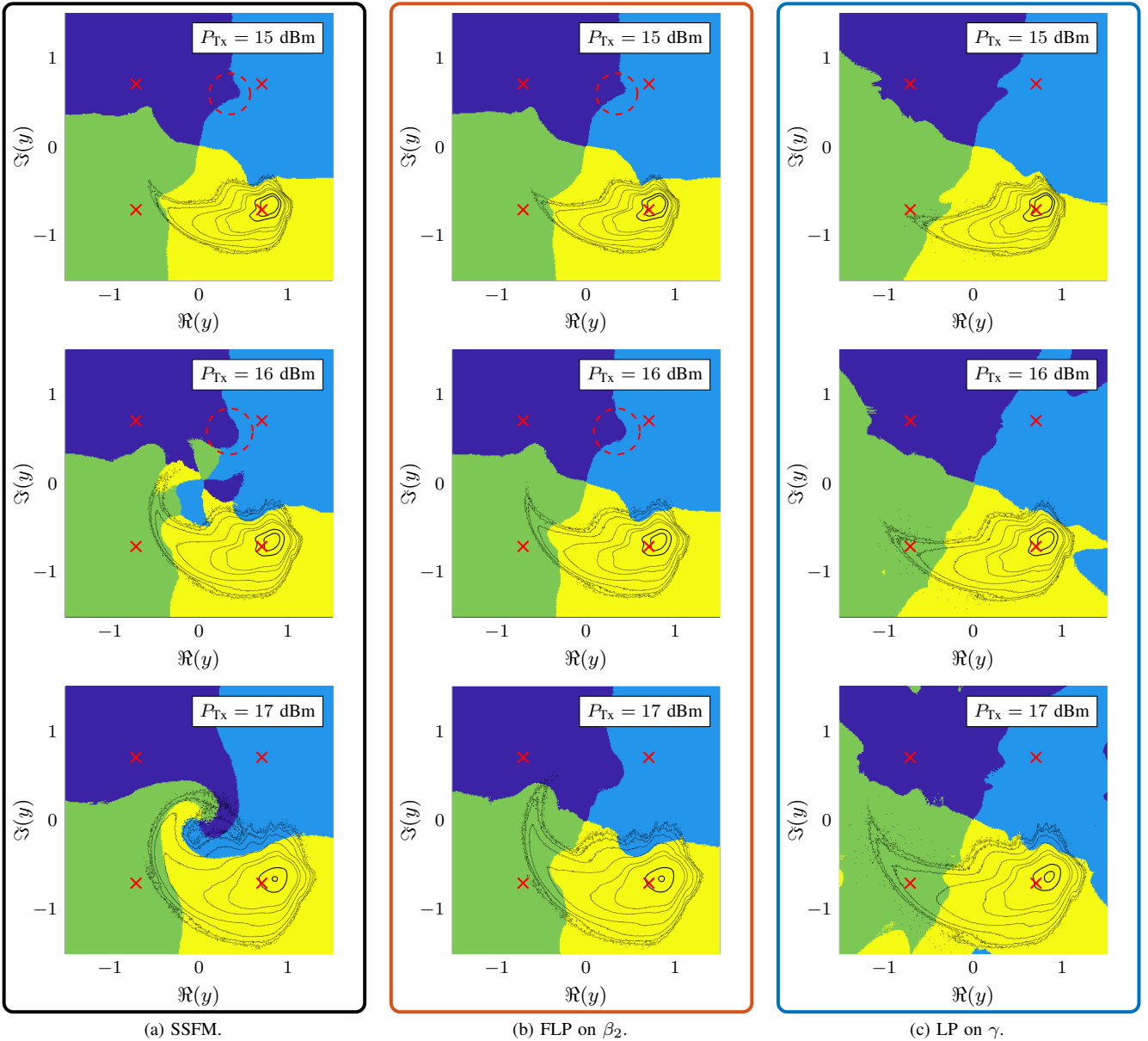

\centering
\subfloat[SSFM.]{\input{figures/constellationSSFM.tikz}
\label{fig:constssfm}}
\hspace{0.75cm}
\subfloat[FLP on $\beta_2$.]{\input{figures/constellationB2.tikz} \label{fig:constb2}}
\hspace{0.75cm}
\subfloat[LP on $\gamma$.]{\input{figures/constellationgamma.tikz}
\label{fig:constgamma}}
\caption{Optimized decision regions obtained with HB detection after the propagation for $P_{\text{Tx}}\in \{15,16,17\}$ dBm of a QPSK constellation based on: (a) SSFM; (b) FLP on $\beta_2$; (c) LP on $\gamma$. Each of the four differently colored areas represents the decision region for a specific constellation point. The respective constellation points are illustrated with red crosses. A contour plot of the histogram for the received samples for the transmitted symbol $(1-j)/\sqrt{2}$ is plotted on top of each subfigure.}
\label{fig:constmodels}
\end{figure*}

The decision region optimization for the symbol detector is based on the histogram of the received complex samples. This histogram-based (HB) detector follows the principle of choosing the most probable transmitted symbol, given that the corresponding received sample falls in certain area of the complex plane. This principle corresponds to the maximum a-posteriori (MAP) rule 
\begin{equation}\label{eq:map}
    \hat{x} = \argmax_{s_m \in \mathcal{S}} \text{Pr}\{X=s_m \ | \ Y=y\}.
\end{equation}
where $X$ and $Y$ are random variables associated with the transmitted symbols and received samples (see Fig.~\ref{fig:ftthsyst}), respectively, and
$s_m \in \mathcal{S}$ where $\mathcal{S} = \{s_1,\cdots,s_M\}$ is the set of constellation points with cardinality $M$ and $1 \le m \le M$.
The MAP rule, as stated in \eqref{eq:map}, is optimal for a memoryless channel, and thus, suboptimal for the optical fiber channel, which includes memory.


To build the HB detector and numerically approximate the rule in \eqref{eq:map}, training symbols are transmitted through the fiber in order to obtain an estimation of the probability distribution. Their respective received samples will fall in a specific bin, which is an small area in the complex plane. For each bin, we count the number $n_{m,b}$ of received samples that fall inside the $b$-th bin and were drawn from the $m$-th constellation point, where $1 \le b \le B$, $1\le m \le M$, and $B$ is the total number of bins. The value of $m=\hat{m}_b$ that maximizes $n_{m,b}$ is considered to be the most probable transmitted constellation point for that bin. If $\max_m\{n_{m,b}\} = 0$, we replace $b$ by the closest bin $b'$ in Euclidean distance such that $\max_m\{n_{m,b'}\} > 0$, and decide $\hat{m}_b = \text{argmax}_m\{n_{m,b'}\}$. After obtaining $\hat{m}_b$ for all possible $b$, the decision regions are defined and every time a received sample $y$ is received in the $b$-th bin, we assume that received symbol $\hat{x} = s_{\hat{m}_b}$ was transmitted. 
For computational reasons, we only consider a subset $\mathcal{A} = \{z \in \mathbb{C} \, : \, |\Re\{z\}|<2,|\Im\{z\}| < 2\}$ of the complex plane, divided into $B=400\times400$ square bins of side 0.01. This region is sufficient to contain virtually all the received samples in the simulation. We simulate approximately $5.4\cdot 10^8$ symbols to obtain histograms that define $n_{m,b}$. The decision regions were optimized for each transmitted power separately.

Fig.~\ref{fig:constmodels} depicts the decision regions obtained by using HB detectors for SSFM, FLP on $\beta_2$, and LP on $\gamma$ at input powers $P_\text{Tx} \in \{15,16,17\}$ dBm. The transmitted constellation is illustrated with red crosses. Fig.~\ref{fig:constmodels} also includes contour plots of the histogram for the received samples when transmitting the constellation point $(1-j)/\sqrt{2}$. The decision regions associated to this constellation point are represented in yellow and contain most of the received samples shown by the contour plots. The samples that fall outside the yellow regions are not classified as $(1-j)/\sqrt{2}$ since there are more received samples originating from another constellation point in those specific bins.

\begin{figure}[!t]
\centering
    
\def\slo{\small}
\def\to{\footnotesize}

\def\xsiz{1.4}
\def\linW{0.8}

\definecolor{mycolor1}{rgb}{0.00000,0.44700,0.74100}%
\definecolor{mycolor2}{rgb}{0.00000,0.49804,0.00000}%
\definecolor{mycolor3}{rgb}{0.92900,0.69400,0.12500}%
\definecolor{mycolor4}{rgb}{0.85098,0.32549,0.09804}%
\definecolor{mycolor5}{rgb}{0.00000,0.44706,0.74118}%
\definecolor{mycolor6}{rgb}{0.0,0.1,0.08}%
\definecolor{mycolor7}{rgb}{0.72,0.3,0.08}%

\begin{tikzpicture}

\begin{axis}[%
width=2.75in,
height=2.45in,
at={(0in,0in)},
scale only axis,
xmin=14,
xmax=20,
xlabel style={font=\color{white!15!black},font=\slo},
xlabel={Input power [dBm]},
ymode=log,
ymin=1e-5,
ymax=1e-1,
yminorticks=true,
ylabel style={font=\color{white!15!black},font=\slo},
ylabel={BER},
axis background/.style={fill=white},
xlabel shift = -0.03in,
ylabel shift = -0.04in,
xmajorgrids,
ymajorgrids,
yminorgrids,
ticklabel style = {font=\to},
 legend columns=1, 
legend style={at={(0.55625,0.00)}, anchor=south west, legend cell align=left, align=left, draw=white!15!black,font=\scriptsize,            /tikz/column 2/.style={
                column sep=5pt,
            }}
]

\addplot [color=mycolor2, line width=0.8pt,mark size=1pt,  mark=*, mark options={fill=white,solid,scale=1.5}]
  table[row sep=crcr]{%
14	0.000233173370361328\\
14.5 6.781578063964844e-04\\
15	0.00161647796630859\\
15.5 0.003310251235962\\
16	0.00596761703491211\\
16.5 0.009758996963501\\
17	0.0150814056396484\\
17.5 0.021917605400085\\
18	0.0305805206298828\\
18.5 0.039522790908813\\
19	0.048736572265625\\
19.5 0.058958053588867\\
20	0.0715880393981934\\
};
\addlegendentry{Min. dist. HB}

\addplot [color=black, line width=0.8pt,mark size=1pt,  mark=*, mark options={fill=white,solid,scale=1.5}]
  table[row sep=crcr]{%
14	0\\
14.5 8.821487426757813e-7\\
15	1.363754272460938e-05\\
15.5 8.339881896972657e-05\\
16	0.000354290008544922\\
16.5 0.001054954528809\\
17	0.00234460830688477\\
17.5 0.004779481887817\\
18	0.00992584228515625\\
18.5 0.018761467933655\\
19	0.0307502746582031\\
19.5 0.045082616806030\\
20	0.0615882873535156\\
};
\addlegendentry{SSFM HB}

\addplot [color=mycolor4,mark size=0.9pt, line width=0.8pt,  mark=square*, mark options={fill=white,solid,scale=1.5}]
  table[row sep=crcr]{%
14	0\\
14.5 1.358985900878906e-06\\
15	1.506805419921875e-05\\
15.5 9.241104125976563e-05\\
16	0.000391960144042969\\
16.5 0.001231479644775\\
17	0.00337076187133789\\
17.5 0.006826090812683\\
18	0.0118002891540527\\
18.5 0.019268894195557\\
19	0.0310568809509277\\
19.5 0.045672154426575\\
20	0.0624332427978516\\
};
\addlegendentry{FLP on $\beta_2$ HB}

\addplot [color=mycolor1,mark size=0.9pt, line width=0.8pt,   mark options={fill=white,solid,scale=1.5}]
  table[row sep=crcr]{%
14	0\\
14.5 6.651878356933594e-06\\
15	7.48634338378906e-05\\
15.5 3.091096878051758e-04\\
16	0.000874519348144531\\
16.5 0.002255058288574\\
17	0.00523138046264648\\
17.5 0.009634184837341\\
18	0.0161385536193848\\
18.5 0.024765443801880\\ 
19	0.0356874465942383\\
19.5 0.049259066581726\\
20	0.0659694671630859\\
};
\addlegendentry{LP on $\gamma$ HB}

\addplot [color=cyan,dashed,line width=1.2pt, mark size=1pt,  mark=*, mark options={fill=white,solid,scale=1.5}]
  table[row sep=crcr]{%
14	2.121925354003906e-05\\
14.5 7.840991020202637e-05\\
15	2.259910106658936e-04\\
15.5 4.850327968597412e-04\\
16	0.001013100147247\\
16.5 0.002014487981796\\
17	0.003731489181519\\
17.5 0.006510853767395\\
18	0.011860758066177\\
18.5 0.020666569471359\\
19	0.032391607761383\\
19.5 0.046337753534317\\
20	0.062056154012680\\
};
\addlegendentry{PW detec. $T=2^{11}$}

\addplot [color=red!50!white,dashed,line width=1.2pt, mark size=1pt,  mark=*, mark options={fill=white,solid,scale=1.5}]
  table[row sep=crcr]{%
14 3.695487976074219e-05\\
14.5 1.268386840820313e-04\\
15 3.372728824615479e-04	\\
15.5 7.349848747253418e-04\\
16 0.001459151506424	\\
16.5 0.002569913864136\\
17 0.004517108201981	\\
17.5 0.007679700851440\\
18 0.013024598360062\\
18.5 0.022254943847656\\
19 0.034256398677826\\
19.5 0.048300087451935\\
20 0.064026892185211\\
};
\addlegendentry{PW detec. $T=2^{10}$}

\draw[thick,dashed] (16,0.000354290008544922) -- (15.5,0.000354290008544922);
\draw[thick,dashed] (16,0.00596761703491211) -- (15.5,0.00596761703491211);

\draw [<->,>=stealth] (15.6,0.000354290008544922)--(15.6,0.00596761703491211);

\node [align=center] at (15.9,0.003) {\scriptsize{$\approx$} \\ [-0.85ex]  \scriptsize{$ \times 17$}};

\draw[thick,dashed] (15,1.38282775878906e-05) -- (15.5,1.38282775878906e-05);
\draw[thick,dashed] (15,7.48634338378906e-05) -- (15.5,7.48634338378906e-05);

\draw [<->,>=stealth] (15.4,1.38282775878906e-05)--(15.4,7.48634338378906e-05);

\node [align=center] at (15.7,3e-5) {\scriptsize{$\approx$} \\ [-0.85ex]  \scriptsize{$ \times 5.4$}};

\node[ellipse,semithick,minimum height=1cm,minimum width=0.65cm,draw] at (17,3.6e-3) (ell1) {};

\draw (ell1) -- (19,0.002);

\draw[thick,dashed] (15, 0.000339984893798828) -- (14.5,0.000339984893798828);
\draw[thick,dashed] (15,	2.259910106658936e-04) -- (14.5, 2.259910106658936e-04);

\draw [<->,>=stealth] (14.6,0.000339984893798828)--(14.6,2.259910106658936e-04);

\node (txt15) [align=center] at (14.3,2.3e-3) {\scriptsize{$\approx$} \\ [-0.85ex]  \scriptsize{$ \times 1.5$}};

\draw[->] (14.6,0.00025) -- (txt15);

\draw[dashed,thick] (14,0.0011) -- (20,0.0011);

\draw[dashed,thick] (14,0.00018) -- (20,0.00018);

\node [align=center] (txtRS239) at (16.65,0.000175) { \scriptsize{FEC limit}\\ [-0.85ex] \scriptsize{$\text{RS}(255,239)$}};

\node [align=center] (txtRS223) at (17.65,0.00105) { \scriptsize{FEC limit}\\ [-0.85ex] \scriptsize{$\text{RS}(255,223)$}};

\node (pwtxt) at (17,4.75e-4) {\scriptsize{$16.5$ dBm}};

\draw[->] (16.5,1.1e-3) -- (pwtxt);

\end{axis}

\begin{axis}[%
width=0.7in,
height=0.7in,
at={(2in,1in)},
scale only axis,
xmin=16.6,
xmax=17.1,
xlabel style={font=\color{white!15!black},font=\slo},
ymode=log,
ymin=1.8e-3,
ymax=5e-3,
xtick=\empty,
ytick=\empty,
ylabel style={font=\color{white!15!black},font=\slo},
axis background/.style={fill=white},
xlabel shift = -0.03in,
ylabel shift = -0.04in,
ticklabel style = {font=\to},
 legend columns=1, 
legend style={at={(0.58,0.03)}, anchor=south west, legend cell align=left, align=left, draw=white!15!black,font=\scriptsize,            /tikz/column 2/.style={
                column sep=5pt,
            }}
]

\addplot [color=black, line width=0.8pt,mark size=1pt,  mark=*, mark options={fill=white,solid,scale=1.5}]
  table[row sep=crcr]{%
16	0.000354290008544922\\
16.5 0.001054954528809\\
17	0.00234460830688477\\
17.5 0.004779481887817\\
18	0.00992584228515625\\
};

\addplot [color=mycolor4,mark size=0.9pt, line width=0.8pt,  mark=square*, mark options={fill=white,solid,scale=1.5}]
  table[row sep=crcr]{%
16	0.000391960144042969\\
16.5 0.001231479644775\\
17	0.00337076187133789\\
17.5 0.006826090812683\\
18	0.0118002891540527\\
};

\addplot [color=mycolor1, mark size=2.5pt, line width=0.8pt,mark=asterisk]
  table[row sep=crcr]{%
16	0.000874519348144531\\
16.5 0.002255058288574\\
17	0.00523138046264648\\
17.5 0.009634184837341\\
18	0.0161385536193848\\
};

\addplot [color=cyan,dashed,line width=1.2pt, mark size=1pt,  mark=*, mark options={fill=white,solid,scale=1.5}]
  table[row sep=crcr]{%
16	0.001013100147247\\
16.5 0.002014487981796\\
17	0.003731489181519\\
17.5 0.006510853767395\\
18	0.011860758066177\\
};

\addplot [color=red!50!white,dashed,line width=1.2pt, mark size=1pt,  mark=*, mark options={fill=white,solid,scale=1.5}]
  table[row sep=crcr]{%
16.5 0.002569913864136\\
17 0.004517108201981	\\
17.5 0.007679700851440\\
18 0.013024598360062\\
};

\draw[thick,dashed] (17,0.00234460830688477) -- (16.825,0.00234460830688477);
\draw[thick,dashed] (17,0.00337076187133789) -- (16.825,0.00337076187133789);

\draw [<->,>=stealth] (16.87,0.00234460830688477)--(16.87,0.00337076187133789);

\node (txt3) [align=center] at (16.7,0.00425) {\scriptsize{$\approx$} \\ [-0.85ex]  \scriptsize{$ \times 1.4$}};

\draw [->] (16.87,0.0027) -- (txt3.south);

\end{axis}

\end{tikzpicture}
\caption{BER versus input power for receivers using HB detectors and PW detectors. The HB dectors were obtained via SSFM, LP on $\gamma$, and FLP on $\beta_2$. In all cases, the channel was simulated in the C-band using the SSFM. Min. dist.: Minimum distance.}
\label{fig:ber-resul}
\end{figure}
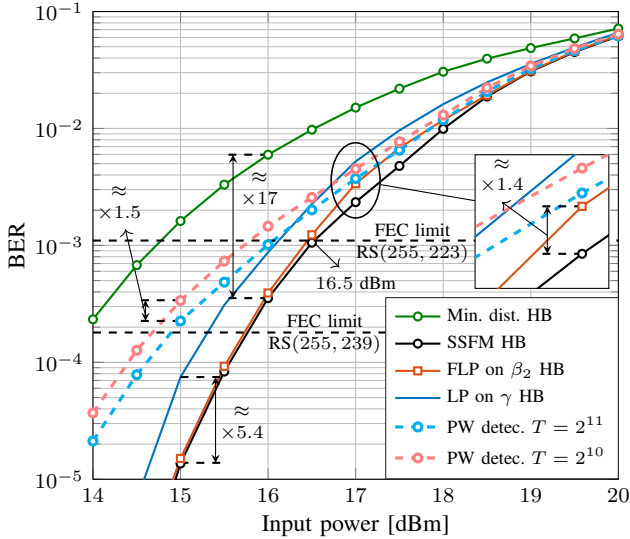

The optimum decision regions are considered as the ones obtained via SSFM in Fig.~\subref*{fig:constssfm}. At 15 dBm, the decision regions differ significantly from the four quadrants of the complex plane that represent the minimum distance decision regions. This difference originates from the nonlinearities, which creates a non-Gaussian distribution of the received samples. Due to the high nonlinear effect present at 17 dBm, the SSFM decision regions and the contour plots follow a spiral shape. 
The decision regions obtained by FLP on $\beta_2$ are shown in Fig.~\subref*{fig:constb2}. At 15 dBm, these decision regions are similar to the SSFM ones for the same $P_{\text{Tx}}$. However, at 17 dBm, these regions do not present a spiral shape, as in Fig.~\subref*{fig:constssfm}. This new behavior can be explained by the contour plots. The curvature of the contour plots in Fig.~\subref*{fig:constb2} is lower than the curvature in Fig.~\subref*{fig:constssfm}, which results in the symbols from the tail of the contour plot falling in bins with a high number of points transmitted from another constellation point. The curvature for the contour plot of symbols obtained via LP on $\gamma$ in Fig.~\subref*{fig:constgamma} is even lower than for FLP on $\beta_2$ in Fig.~\subref*{fig:constb2}. This fact results in decision regions with less accentuated curves in center of Fig.~\subref*{fig:constgamma} for each $P_{\text{Tx}}$. At $15$ dBm, the LP on $\gamma$ decision regions already differ significantly from the SSFM ones. For example, at both $15$ and $16$ dBm, a small lobe (red dashed circle) present in both SSFM and FLP on $\beta_2$ decision regions is not present in the LP on $\gamma$ ones. The shape of the contour plots indicate that the decision regions obtained with FLP on $\beta_2$ might perform closer to the SSFM decision regions than the ones obtained with LP on $\gamma$, as will be discussed next.

The HB detectors obtained using LP on $\gamma$, FLP on $\beta_2$, and SSFM are compared in a system whose fiber propagation is modeled by the SSFM. The results are shown in Fig.~\ref{fig:ber-resul}, where the BER is evaluated for different launch powers using the obtained HB detectors and PW detectors. The latter are obtained using the SSFM for fiber propagation. As shown in Fig.~\ref{fig:ber-resul}, the SSFM HB detector (black curve) shows the lowest BER for the displayed input powers. The minimum distance HB detector gives the worst performance since it assumes a Gaussian distribution of the received samples. This detector is obtained by minimum Euclidean distance from the received samples to the possible transmitted symbols. Replacing the minimum distance HB detector by the SSFM one at 16 dBm reduces the BER approximately 17 times (from $6.0\cdot 10^{-3}$ to $3.5\cdot 10^{-4}$). At 15 dBm, the BER for FLP on $\beta_2$ is 5.4 times lower than for LP on $\gamma$. As expected, results for FLP on $\beta_2$ are closer to the SSFM results than LP on $\gamma$. The highest gap between the SSFM and FLP on $\beta_2$ occurs at 17 dBm, where the BER for FLP on $\beta_2$ is 1.4 times higher than the BER for SSFM.


For the PW detector \cite{karim2019}, instead of defining bins in the complex plane, a certain amount ($T$) of the $N$ received samples is used as training symbols, denoted by $y_t$, $1\le t \le T$. The remaining received samples are used for testing, denoted by $y_k$, where $T+1\le k \le N$. The received symbol $\hat{x}$ depends on the distance between $y_k$ and all the training symbols $y_t$ in the following way:
for every received sample $y_k$, we identify a set of nearby training samples by calculating the set of time indices $\mathcal{T}_s = \{t=1,\ldots,T: x_t = s, |y_t-y_k| \le R \}$ for every $s \in \mathcal{S}$ and for a given radius $R>0$. For each transmitted symbol $s$, $\sum_{t \in \mathcal{T}_s} (1/|y_t-y_k|)$ is calculated, and the symbol with the highest sum is taken as the received symbol $\hat{x}_k$.
In our simulations, we considered two scenarios: $T=2^{10}$ and $T=2^{11}$, both with $N=2^{16}$. Each scenario was repeated 256 times and the resulting BER averaged over the repetitions. The radius $R$ was optimized for each transmitted sequence with a grid search, as illustrated in \cite[Fig.~3]{karim2019}.

As depicted in Fig.~\ref{fig:ber-resul}, the PW detector BER for $2^{11}$ training symbols converges to the BER of FLP on $\beta_2$ at input powers higher than 17 dBm. At input powers lower than 16 dBm, the PW BER for $2^{11}$ training symbols is even higher than the BER for LP on $\gamma$. This result shows that the HB detector can perform close to recently proposed methods in the literature. At 15 dBm, the penalty obtained for reducing the number of training symbols from $2^{11}$ to $2^{10}$ is an increase in BER of approximately 1.5 times. The PW detector performance could be further improved by increasing $T$, at the cost of increased complexity at the receiver.

\subsection{Achievable Information Rates}\label{sc:achivrat}

FEC is present in modern PON systems to improve system performance \cite{ITUXG9873}. In Fig.~\ref{fig:ber-resul}, pre-FEC BER thresholds are shown for two Reed-Solomon (RS) codes \cite{ReedSolomon60}. The considered codes in Fig.~\ref{fig:ber-resul} are $\text{RS}(n,k)$ with $k=239,223$ and $n=255$, where $k$ and $n$ are the information and codeword lengths, resp. These two codes are typical low-complexity RS codes used in PONs \cite{Schmalen2013,ITUXG9873} and have a code rate of $0.93$ and $0.87$, respectively. 

Along with RS codes, also stronger FEC codes such as low-density parity-check codes or staircase codes have been proposed for PONs in the literature \cite{LatencyNokia,FEC2550100Nokia,Teixeira:20}. In this section, we evaluate the models for PON systems with hard-decision (HD) FEC in terms of achievable information rates (AIRs) \cite{Alvarado2018AIR}. We consider the simple RS codes described above and a theoretical HD limit for the AIR. The latter is close to the performance of strong HD FEC codes \cite[Fig.\ 8]{Smith:12}. The AIRs are obtained by closed-form expressions based on pre-FEC BER.



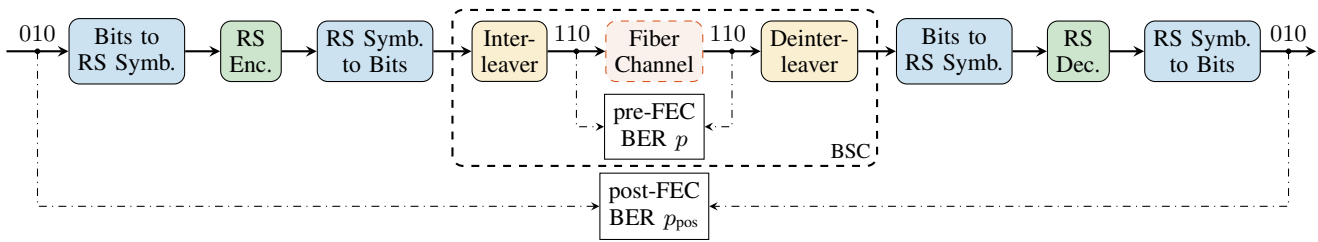
\begin{figure*}[!t]
\centering
\setlength{\hsep}{1cm}

\tikzstyle{Cir} = [draw, circle, 
minimum size=1.5em]

\begin{tikzpicture}
\definecolor{mycolor1}{rgb}{0.00000,0.44706,0.74118}%
\definecolor{mycolor2}{rgb}{0.00000,0.49804,0.00000}%
\definecolor{mycolor3}{rgb}{0.92900,0.69400,0.12500}%
\definecolor{mycolor4}{rgb}{0.85098,0.32549,0.09804}%
\definecolor{mycolor5}{rgb}{0.39,0.19,0.0625};

\def\sca{2.75}
\def\Cwd{0.03}

\def\df{0.6*\sca}
\def\denctwo{0.6*\sca}
\def\dcotwo{0.6*\sca}

\def\dint{0.65*\sca}

\def\dfib{0.7*\sca}
\def\dinttwo{0.75*\sca}

\def\dspl{0.6*\sca}
\def\dspltwo{0.7*\sca}

\def\douttwo{0.55*\sca}

\def\symbI{0.3*\sca}
\def\symbO{0.3*\sca}

\def\vjump{-0.365*\sca}
\def\vjumptwo{-0.75*\sca}

\coordinate (CO) at (0,0);
\coordinate [right=\df of CO] (Fiber); 

\coordinate [right=\dspl of Fiber] (Spl);
\coordinate [right=\dint of Spl] (INTC);

\coordinate [right=\dfib of INTC] (FibSysC);
\coordinate [right=\dinttwo of FibSysC] (INTC2);
\coordinate [right=\dspltwo of INTC2] (Spl2);
\coordinate [right=\denctwo of Spl2] (Fiber2);
\coordinate [right=\dcotwo of Fiber2] (CO2);

\coordinate [right=\douttwo of CO2] (out2);


\node[align=center] (NCO) [rectangle,draw,rounded corners,fill=mycolor1,fill opacity=0.2,text opacity = 1] at (CO) {\small{Bits to}\\[-0.2em]\small{RS Symb.}};

\node[align=center] (RSENC)  [rectangle,draw,rounded corners,fill=mycolor2,fill opacity=0.2,text opacity = 1] at (Fiber) {\small{RS}\\[-0.2em]\small{Enc.}};

\node[align=center] (INTB)  [rectangle,draw,rounded corners,fill=mycolor3,fill opacity=0.2,text opacity = 1] at (INTC) {\small{Inter-}\\[-0.2em]\small{leaver}};

\draw[thick,->,>=stealth, align=center] (NCO.east)--  (RSENC.west);

\node[align=center] (NRx) [rectangle,draw,rounded corners,fill=mycolor1,fill opacity=0.2,text opacity = 1] at (Spl) {\small{RS Symb.}\\[-0.2em]\small{to Bits}};
\coordinate (ini) at ($(NCO.west)+(-\symbI,0)$);

\draw[thick,->,>=stealth] (ini)-- node[above] {\small{$010$}} (NCO);

\coordinate (out) at ($(INTB.east)+(\symbO,0)$);

\draw[thick,->,>=stealth, align=center] (RSENC.east)-- (NRx.west);
\draw[thick,->,>=stealth, align=center] (NRx.east)-- (INTB.west);

\coordinate (ini2) at ($(ini)+(0,\vjump)$);


\node[align=center,dashed,mycolor4,text=black] (FibSysB) [rectangle,draw,rounded corners,fill=mycolor4!10!white,fill opacity=0.8,text opacity = 1] at (FibSysC) {\small{Fiber}\\[-0.2em]\small{Channel}};

\node[align=center] (NCO2) [rectangle,draw,rounded corners,fill=mycolor1,fill opacity=0.2,text opacity = 1] at (CO2) {\small{RS Symb.}\\[-0.2em]\small{to Bits}};

\node[align=center] (RSENC2)  [rectangle,draw,rounded corners,fill=mycolor2,fill opacity=0.2,text opacity = 1] at (Fiber2) {\small{RS}\\[-0.2em]\small{Dec.}};

\node[align=center] (NRx2) [rectangle,draw,rounded corners,fill=mycolor1,fill opacity=0.2,text opacity = 1] at (Spl2) {\small{Bits to}\\[-0.2em]\small{RS Symb.}};

\node[align=center] (INTB2) [rectangle,draw,rounded corners,fill=mycolor3,fill opacity=0.2,text opacity = 1] at (INTC2) {\small{Deinter-}\\[-0.2em]\small{leaver}};



\draw[thick,->,>=stealth, align=center] (INTB.east)-- node[above] {\small{$110$}} (FibSysB.west);
\draw[thick,->,>=stealth, align=center] (FibSysB.east)-- node[above] {\small{$110$}} (INTB2.west);

\draw[thick,<-,>=stealth, align=center] (NCO2.west)-- (RSENC2.east);
\draw[thick,<-,>=stealth, align=center] (RSENC2.west)-- (NRx2.east);
\draw[thick,<-,>=stealth, align=center] (NRx2.west)-- (INTB2.east);
\draw[thick,<-,>=stealth] (out2)-- node[above] {\small{$010$}} (NCO2);

\coordinate (IBE) at (INTB.east);
\coordinate (FBW) at (FibSysB.west);

\coordinate (IB2W) at (INTB2.west);
\coordinate (FBE) at (FibSysB.east);

\coordinate (preB1UP) at ($(IBE)!0.5!(FBW)$);
\coordinate (preB2UP) at ($(IB2W)!0.5!(FBE)$);

\coordinate (preFECC) at ($(FibSysB)+(0,\vjump)$);

\node[align=center] (preFECB) [rectangle,draw,fill=white,fill opacity=0.2,text opacity = 1] at (preFECC) {\small{pre-FEC} \\ [-0.2em] \small{BER $p$}};

\draw[->,>=stealth,dash dot] (preB1UP) |- (preFECB.west);
\draw[->,>=stealth,dash dot] (preB2UP) |- (preFECB.east);

\coordinate (BRS1W) at (NCO.west);
\coordinate (RSB2E) at (NCO2.east);

\coordinate (posB1UP) at ($(ini)!0.5!(BRS1W)$);
\coordinate (posB2UP) at ($(RSB2E)!0.5!(out2)$);

\coordinate (posFECC) at ($(FibSysB)+(0,\vjumptwo)$);

\node[align=center] (posFECB) [rectangle,draw,fill=white,fill opacity=0.2,text opacity = 1] at (posFECC) {\small{post-FEC} \\ [-0.2em] \small{BER $p_{\text{pos}}$}};

\draw[->,>=stealth,dash dot] (posB1UP) |- (posFECB.west);
\draw[->,>=stealth,dash dot] (posB2UP) |- (posFECB.east);

\draw[fill=black] (preB1UP) circle (\Cwd); 
\draw[fill=black] (preB2UP) circle (\Cwd); 

\draw[fill=black] (posB1UP) circle (\Cwd); 
\draw[fill=black] (posB2UP) circle (\Cwd); 




\coordinate (BSC1) at (INTB2.east);
\coordinate (BSC2) at (INTB.west);
\coordinate (BSC3) at (NRx.east);
\coordinate (BSC4) at (NRx2.west);

\coordinate (BSCL) at ($(BSC2)!0.5!(BSC3)$);
\coordinate (BSCR) at ($(BSC1)!0.5!(BSC4)$);

\def\downbsc{1.525}
\def\upbsc{0.55}

\draw[dashed,rounded corners, color=black,fill=none,thick] ($(BSCL)+(0,-\downbsc)$) rectangle ($(BSCR)+(0,\upbsc)$);

\node at ($(BSCR)+(-0.35,-1.325)$) {\footnotesize{BSC}};

\end{tikzpicture}
\caption{Encoding/decoding procedures used in this paper for RS codes. The fiber channel block include all components from Fig.~\ref{fig:ftthsyst}. The post-FEC BER $p_{\text{pos}}$ is estimated from the pre-FEC BER $p$ using \eqref{eq:rsposfec}.
Enc.: encoder; Dec.: decoder; BSC: binary symmetric channel.}
\label{fig:enc-dec-RS}
\end{figure*}


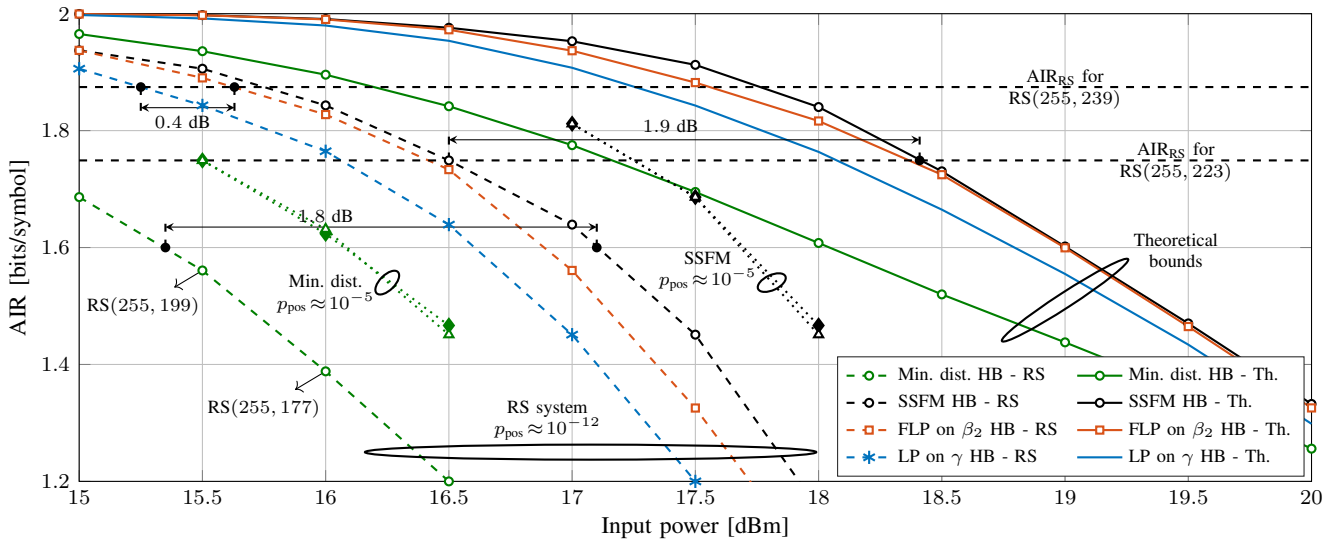
\begin{figure*}[!t]
\centering
    
\def\slo{\small}
\def\to{\footnotesize}

\def\xsiz{1.4}
\def\linW{0.8}

\definecolor{mycolor1}{rgb}{0.00000,0.44700,0.74100}%
\definecolor{mycolor2}{rgb}{0.00000,0.49804,0.00000}%
\definecolor{mycolor3}{rgb}{0.92900,0.69400,0.12500}%
\definecolor{mycolor4}{rgb}{0.85098,0.32549,0.09804}%
\definecolor{mycolor5}{rgb}{0.00000,0.44706,0.74118}%
\definecolor{mycolor6}{rgb}{0.0,0.1,0.08}%
\definecolor{mycolor7}{rgb}{0.72,0.3,0.08}%

\begin{tikzpicture}

\begin{axis}[%
width=6.45in,
height=2.45in,
at={(0in,0in)},
scale only axis,
xmin=15,
xmax=20,
xlabel style={font=\color{white!15!black},font=\slo},
xlabel={Input power [dBm]},
ymin=1.2,
ymax=2,
yminorticks=true,
ylabel style={font=\color{white!15!black},font=\slo},
ylabel={AIR [bits/symbol]},
axis background/.style={fill=white},
xlabel shift = -0.03in,
ylabel shift = -0.04in,
xmajorgrids,
ymajorgrids,
yminorgrids,
ticklabel style = {font=\to},
 legend columns=2, 
legend style={at={(0.615,0.01)}, anchor=south west, legend cell align=left, align=left, draw=white!15!black,font=\scriptsize,            /tikz/column 2/.style={
                column sep=5pt,
            }}
]

\addplot [color=mycolor2, line width=0.8pt,mark size=1pt,  mark=*, dashed, mark options={fill=white,solid,scale=1.5}]
  table[row sep=crcr]{%
15	1.68627450980392\\
15.5	1.56078431372549\\
16	1.38823529411765\\
16.5	1.2\\
17	0.949019607843137\\
17.5	0.682352941176471\\
18	0.368627450980392\\
18.5	0.0392156862745098\\
19	-0.00784313725490196\\
19.5	-0.00784313725490196\\
20	-0.00784313725490196\\
};
\addlegendentry{Min. dist. HB - RS}

\addplot [color=mycolor2, line width=0.8pt,mark size=1pt,  mark=*, mark options={fill=white,solid,scale=1.5}]
  table[row sep=crcr]{%
15	1.96523759762467\\
15.5	1.93584270088144\\
16	1.89564441988541\\
16.5	1.84167371451758\\
17	1.77508791849469\\
17.5	1.69502044011408\\
18	1.60778662114575\\
18.5	1.51985505533876\\
19	1.43782861280052\\
19.5	1.35339084603274\\
20	1.25598414126971\\
};
\addlegendentry{Min. dist. HB - Th.}

\addplot [color=black,dashed,line width=0.8pt,mark size=1pt,  mark=*, mark options={fill=white,solid,scale=1.5}]
  table[row sep=crcr]{%
15	1.93725490196078\\
15.5	1.90588235294118\\
16	1.84313725490196\\
16.5	1.74901960784314\\
17	1.63921568627451\\
17.5	1.45098039215686\\
18	1.15294117647059\\
18.5	0.713725490196078\\
19	0.180392156862745\\
19.5	-0.00784313725490196\\
20	-0.00784313725490196\\
};
\addlegendentry{SSFM HB - RS}

\addplot [color=black,line width=0.8pt,mark size=1pt,  mark=*, mark options={fill=white,solid,scale=1.5}]
  table[row sep=crcr]{%
15	1.99949461480735\\
15.5	1.99750372224763\\
16	1.99090242109935\\
16.5	1.97594883980879\\
17	1.95285712718067\\
17.5	1.91246203279565\\
18	1.84031948723768\\
18.5	1.73036058080879\\
19	1.60176085999618\\
19.5	1.47018145197875\\
20	1.33243574373\\
};
\addlegendentry{SSFM HB - Th.}

\addplot [color=mycolor4,mark size=0.9pt, line width=0.8pt,  mark=square*,dashed, mark options={fill=white,solid,scale=1.5}]
  table[row sep=crcr]{%
15	1.93725490196078\\
15.5	1.89019607843137\\
16	1.82745098039216\\
16.5	1.73333333333333\\
17	1.56078431372549\\
17.5	1.32549019607843\\
18	1.04313725490196\\
18.5	0.682352941176471\\
19	0.180392156862745\\
19.5	-0.00784313725490196\\
20	-0.00784313725490196\\
};
\addlegendentry{FLP on $\beta_2$ HB - RS}

\addplot [color=mycolor4,mark size=0.9pt, line width=0.8pt,  mark=square*, mark options={fill=white,solid,scale=1.5}]
  table[row sep=crcr]{%
15	1.99943361732362\\
15.5	1.99729381318006\\
16	1.99021641352907\\
16.5	1.9725435543667\\
17	1.93665584244588\\
17.5	1.88216522675512\\
18	1.81643173182258\\
18.5	1.72460271818836\\
19	1.59952973564887\\
19.5	1.46483874196685\\
20	1.32566430998153\\
};
\addlegendentry{FLP on $\beta_2$ HB - Th.}

\addplot[color=mycolor1, dashed, mark size=2.5pt, line width=0.8pt,mark=asterisk,mark options={solid}]
  table[row sep=crcr]{%
15	1.90588235294118\\
15.5	1.84313725490196\\
16	1.76470588235294\\
16.5	1.63921568627451\\
17	1.45098039215686\\
17.5	1.2\\
18	0.886274509803922\\
18.5	0.494117647058824\\
19	0.0392156862745098\\
19.5	-0.00784313725490196\\
20	-0.00784313725490196\\
};
\addlegendentry{LP on $\gamma$ HB - RS}

\addplot [color=mycolor1, mark size=2.5pt, line width=0.8pt]
  table[row sep=crcr]{%
15	1.99773245719239\\
15.5	1.99189838138497\\
16	1.9797600615541\\
16.5	1.95360426352377\\
17	1.90749612097442\\
17.5	1.84286998661538\\
18	1.76361305286345\\
18.5	1.66483796520577\\
19	1.55483359289513\\
19.5	1.43391587982273\\
20	1.29850709616834\\
};
\addlegendentry{LP on $\gamma$ HB - Th.}

\addplot[color=mycolor2, dotted, mark size=2.5pt, line width=0.8pt,mark=diamond*,mark options={solid}]
  table[row sep=crcr]{%
15.5	1.74901960784314\\
16	1.62352941176471\\
16.5	1.46666666666667\\
};

\addplot[color=mycolor2, dotted, mark size=2pt, line width=0.8pt,mark=triangle*,mark options={solid,fill=white}]
  table[row sep=crcr]{%
15.5	1.74901960784314\\
16	1.63137254901961\\
16.5	1.45098039215686\\
};

\addplot[color=black, dotted, mark size=2.5pt, line width=0.8pt,mark=diamond*,mark options={solid}]
  table[row sep=crcr]{%
17	1.81176470588235\\
17.5	1.68627450980392\\
18	1.46666666666667\\
};

\addplot[color=black, dotted, mark size=2pt, line width=0.8pt,mark=triangle*,mark options={solid,fill=white}]
  table[row sep=crcr]{%
17	1.81176470588235\\
17.5	1.68627450980392\\
18	1.45098039215686\\
};

\draw[dashed,thick] (15,2*223/255) -- (20,2*223/255);

\draw[dashed,thick] (15,2*239/255) -- (20,2*239/255);

\draw[thick] (17.075,1.25) ellipse (3cm and 0.1cm);
\node[align=center] at (16.9,1.305)  {\scriptsize{RS system} \\ [-0.85ex] \scriptsize{$p_{\text{pos}}\!\approx\!10^{-12}$}};

\draw[thick,rotate around={33:(19,1.51)}] (19,1.51) ellipse (1cm and 0.1cm);
\node[align=center] at (19.45,1.595)  {\scriptsize{Theoretical} \\ [-0.85ex] \scriptsize{bounds}};

\node [align=center] (txtRS239) at (19,2*239/255-0.75/255) { \scriptsize{$\text{AIR}_{\text{RS}}$ for}\\ [-0.85ex] \scriptsize{$\text{RS}(255,239)$}};

\node [align=center] (txtRS223) at (19.45,2*223/255-0.75/255) { \scriptsize{$\text{AIR}_{\text{RS}}$ for}\\ [-0.85ex] \scriptsize{$\text{RS}(255,223)$}};

\draw[thick,dashed] (15.35,1.6) -- (15.35,1.65);
\draw[thick,dashed] (17.1,1.6) -- (17.1,1.65);

\draw [<->,>=stealth] (15.35,1.635)--(17.1,1.635);

\node (gap102) [align=center] at (16,1.655) {\scriptsize{$1.8$ dB}};

\draw[thick,dashed] (15.25,2*239/255) -- (15.25,2*232.1/255);
\draw[thick,dashed] (15.63,2*239/255) -- (15.63,2*232.1/255);

\draw [<->,>=stealth] (15.25,2*234.5/255)--(15.63,2*234.5/255);

\node (gap102) [align=center] at (15.42,2*231.5/255) {\scriptsize{$0.4$ dB}};

\node [inner sep=1.3,circle,fill=black] at (15.25,2*239/255){};

\node [inner sep=1.3,circle,fill=black] at (15.63,2*239/255){};

\node [inner sep=1.3,circle,fill=black] at (15.35,1.6){};

\node [inner sep=1.3,circle,fill=black] at (17.1,1.6){};

\node [inner sep=1.3,circle,fill=black] at (18.41,2*223/255){};

\draw[thick,dashed] (16.5,2*223/255) -- (16.5,2*229/255);
\draw[thick,dashed] (18.41,2*223/255) -- (18.41,2*229/255);

\draw [<->,>=stealth] (18.41,2*227.5/255)--(16.5,2*227.5/255);

\node (gap102) [align=center] at (17.4,2*230.5/255) {\scriptsize{$1.9$ dB}};



\draw[thick,rotate around={45:(16.25,1.54)}] (16.25,1.54) ellipse (0.2cm and 0.1cm);
\node[align=center] at (16,1.52)  {\scriptsize{Min. dist.} \\ [-0.85ex] \scriptsize{$p_{\text{pos}}\!\approx\!10^{-5}$}};

\draw[thick,rotate around={25:(17.81,1.54)}] (17.81,1.54) ellipse (0.2cm and 0.1cm);
\node[align=center] at (17.55,1.56)  {\scriptsize{SSFM} \\ [-0.85ex] \scriptsize{$p_{\text{pos}}\!\approx\!10^{-5}$}};

\node (txRS1) at (15.75,1.325) {\scriptsize{$\text{RS}(255,177)$}};

\draw [->] (16,1.38823529411765) -- (txRS1);

\node (txRS2) at (15.26,1.5) {\scriptsize{$\text{RS}(255,199)$}};

\draw [->] (15.5,1.56078431372549) -- (txRS2);



\end{axis}

\end{tikzpicture}%
\caption{Comparison of AIRs when using the decision regions obtained in Sec.~\ref{sc:decreg}. The AIRs were obtained  using \eqref{eq:airrs} and \eqref{eq:rsposfec} at $p_{\text{pos}}\approx10^{-12}$ for the RS system and \eqref{eq:airth} for the theoretical bound. The dotted lines represent the comparison between $\text{AIR}_\text{RS}$ when approximating $p_\text{pos}$ by \eqref{eq:rsposfec} (white triangles) and $\text{AIR}_\text{RS}$ when simulating the system in Fig.~\ref{fig:enc-dec-RS} (filled diamonds), both for $p_{\text{pos}}\approx10^{-5}$. Th.: theoretical.}
\label{fig:air-RS}
\end{figure*}

We consider a family of RS codes $\text{RS}(n,k)$ with multiple coding rates, where $k$ is varied to obtain different code rates. We use a fixed codeword length ($n=255$ symbols) in order to constrain the code complexity. For every launch power, $k$ is determined by finding the highest $k$ such that the post-FEC BER falls below a certain threshold. Following \cite[Table\ IV.2]{ITU2016}, we use $10^{-12}$ as post-FEC BER threshold. 

The post-FEC BER $p_{\text{pos}}$ can be approximated by substituting the pre-FEC BER $p$ from the system in Fig.~\ref{fig:ftthsyst} in the analytical expression for binary symmetric channels (BSCs) and bounded-distance decoders \cite{ITUXG9751-equation}
\begin{equation}\label{eq:rsposfec}
    p_{\text{pos}} \approx \frac{1}{n}\sum_{r=t+1}^{n}  \left(\dfrac{ p}{p_s}r+\dfrac{1}{2(t-1)!}\right)  {n \choose r}p_s^r(1-p_s)^{n-r},
\end{equation}
where $t=\lfloor (n- k)/2\rfloor$ is the RS error-correction capability, $p_s=1-(1-p)^m$ is the (RS) symbol error probability, and $m=\lceil\log_2(n+1)\rceil$ is the number of bits per symbol. To improve the total computation time, a binary search algorithm on $k$ over all integers between 1 and 253 is performed. After finding $k$, the AIR for the RS system ($\text{AIR}_{\text{RS}}$) is determined by
\begin{equation}\label{eq:airrs}
    \text{AIR}_{\text{RS}} = \log_2(M) \dfrac{k}{n} = \dfrac{2}{255}k,
\end{equation}
since $M=4$ for QPSK. We call \eqref{eq:airrs} an achievable information rate since we assume that $p_{\text{pos}}<10^{-12}$ can be considered virtually error-free for the system considered in this paper.

Since the expression in \eqref{eq:rsposfec} is valid for BSCs, we need to modify the system in Fig.~\ref{fig:ftthsyst} to fulfill that property. The fiber channel in Fig.~\ref{fig:ftthsyst} presents memory due to the interaction of dispersion and nonlinearities. Therefore, in our simulations we included a bit interleaver and a bit deinterleaver so that the fiber channel in Fig.~\ref{fig:ftthsyst} is well-approximated by a BSC. The resulting system, together with the RS encoding and decoding blocks, can be seen in Fig.~\ref{fig:enc-dec-RS}. The system in Fig.~\ref{fig:enc-dec-RS} was only simulated for $p_{\text{pos}} \approx 10^{-5}$ and was used to validate the results of \eqref{eq:rsposfec}. After this validation, \eqref{eq:rsposfec} was used instead of simulating the system in Fig.~\ref{fig:enc-dec-RS}.

Fig.~\ref{fig:air-RS} depicts the AIRs for the RS system using \eqref{eq:rsposfec}. Before analyzing the results for $p_{\text{pos}}\approx10^{-12}$, we validate \eqref{eq:rsposfec} at $p_{\text{pos}}\approx10^{-5}$ to determine if the system in Fig.~\ref{fig:enc-dec-RS} can be well-approximated by a BSC. This validation is done assuming a minimum distance detector and is shown by the dotted lines in Fig.~\ref{fig:air-RS}. The results for the RS simulations of the exact system in Fig.~\ref{fig:enc-dec-RS} at $p_{\text{pos}}\approx10^{-5}$ are shown with diamonds. An almost perfect overlap with the results from \eqref{eq:rsposfec} with the same $p_{\text{pos}}$ (shown with triangles) is observed. The agreement between these curves suggest that the system in Fig.~\ref{fig:enc-dec-RS} can be approximated by a BSC. Therefore, from now on only \eqref{eq:rsposfec} is used for the RS systems at $p_{\text{pos}}\approx10^{-12}$. 

We start by comparing the dashed lines in Fig.~\ref{fig:air-RS}, which represent the $\text{AIR}_{\text{RS}}$ results in \eqref{eq:airrs}. As shown in Fig.~\ref{fig:air-RS}, using the decision regions obtained by SSFM in a RS system can provide a gain of approximately $1.8$ dB for a rate of $1.6$ bits/symbol in terms of nonlinear tolerance. The crossing point between the rate of $\text{RS}(255,223)$ with $\text{AIR}_{\text{RS}}$ for the SSFM decision regions is at $16.5$ dBm, which closely matches with the crossing point with BER in Fig.~\ref{fig:ber-resul}. When comparing the models, the histogram-based detector obtained using FLP on $\beta_2$ outperforms the LP on $\gamma$ one throughout the considered power range, analogously to Fig.~\ref{fig:ber-resul}. Specifically for $\text{RS}(255,239)$, FLP on $\beta_2$ outperforms LP on $\gamma$ by approximately $0.4$ dB.

The $\text{AIR}_{\text{RS}}$ is also compared with a theoretical bound on hard-decision bit-wise AIRs for independent, identically distributed bit errors. The theoretical AIR ($\text{AIR}_{\text{TH}}$) used in this paper is defined as \cite{Fehenberger:15} 
\begin{equation}\label{eq:airth}
    \text{AIR}_{\text{TH}} = \log_2(M)\left( 1- \text{H}_\text{b}(p)\right),
\end{equation}
where $\text{H}_\text{b}(p)= -p\log_2(p)-(1-p)\log_2(1-p)$ is the binary entropy function for the given pre-FEC BER $p$ in the system of Fig.~\ref{fig:ftthsyst}. The $\text{AIR}_{\text{TH}}$ from \eqref{eq:airth} can be approached by strong FEC codes such as staircase codes, as reported in \cite[Fig.\ 8]{Smith:12}.

As shown in Fig.~\ref{fig:air-RS}, the theoretical bounds (solid lines) from \eqref{eq:airth} show significant gains over the $\text{AIR}_{\text{RS}}$ results. These gains show that, by using codes more complex than $\text{RS}(255,k)$, higher code rates can be achieved or the input power can be improved for a specific rate. For the same code rate as in $\text{RS}(255,223)$, the input power for the theoretical bound on the SSFM decision regions is approximately $1.9$ dB higher than the one for the RS system. However, the complexity and latency of codes that perform close to the theoretical bound should be carefully analysed for a PON system design. The results for the theoretical bounds in in Fig.~\ref{fig:air-RS} also show that the histogram-based detector based on the FLP on $\beta_2$ outperforms the one based on LP on $\gamma$.

\section{Conclusions}\label{sc:conc}

In this paper we presented a novel model for optical fiber transmission and evaluated its performance for a passive optical network system. The proposed model was derived as an improved version of the regular perturbation on the GVD parameter model. The improvement was obtained by applying frequency logarithmic perturbation on the GVD parameter. Both regular and frequency logarithmic perturbation on the GVD parameter models are suitable in the weakly dispersive and highly nonlinear regime, whereas the frequency logarithmic perturbation is able to surpass the limitations of the regular perturbation.

Apart from the regular perturbation on the GVD parameter, the proposed model was compared with two other models present in the literature: regular and logarithmic perturbation on the Kerr nonlinear coefficient.
For a fixed normalized squared deviation of $0.1\%$, the proposed model was accurate at $1.5$ dB higher input powers compared to logarithmic perturbation on the Kerr nonlinear coefficient.
Both frequency logarithmic and regular perturbation on the GVD parameter exhibit the highest convergence rate to the split-step Fourier method results when reducing the dispersion effect. The proposed model also proved to be more suitable for symbol and bit detection, with and without FEC.

Possible extensions of this work are higher-order logarithmic perturbation models, perturbation on the GVD parameter for dual-polarization systems, and nonlinearity-compensation techniques based on the proposed model.

\section*{Acknowledgments}

The authors would like to thank Dr. Tobias Fehenberger (ADVA Optical Networking) for fruitful discussions about the regular perturbation on $\gamma$. We would also like to thank Dr. Nicola Calabretta (Eindhoven University of Technology) and Dr. Domaniç Lavery (Infinera) for insightful discussions about passive optical networks.

\ifCLASSOPTIONcaptionsoff
  \newpage
\fi

\balance

\bibliographystyle{IEEEtran}
\bibliography{IEEEabrv,Bibliography}

\vfill

\end{document}